\documentclass[journal,comsoc]{IEEEtran}
\usepackage{marvosym}
\usepackage{amsmath}
\usepackage{mathtools}

\usepackage{subfig}
\usepackage{siunitx}
\usepackage{verbatim}

\usepackage{amssymb}
\usepackage{booktabs}% http://ctan.org/pkg/booktabs
\usepackage{tabularx}
\usepackage[margin=2cm]{geometry}
\usepackage{tikz}
\usetikzlibrary{automata, positioning}
\usepackage{pgfplots}
\usepackage{graphicx}
\usepackage{epsfig}
\usepackage{epstopdf}
\usepackage{color}
\pgfplotsset{compat=1.11}
\usepackage[T1]{fontenc}
\usepackage{epstopdf}
\usetikzlibrary{calc}
\usepackage{float}
\usepackage{tikz}
\usetikzlibrary{calc}
\usetikzlibrary{shapes.geometric, arrows}
\tikzstyle{startstop} = [ellipse, text centered, draw=black, semithick]
\tikzstyle{io} = [trapezium, trapezium left angle=70, trapezium right angle=110, minimum width=0.3cm, minimum height=0.3cm, text centered, draw=black, semithick]
\tikzstyle{process} = [rectangle, rounded corners, minimum width=1cm, minimum height=0cm, text centered,    draw=black, semithick]
\tikzstyle{decision} = [diamond, minimum width=1cm, minimum height=0.5cm, text centered,  draw=black, semithick]
\tikzstyle{arrow} = [semithick,->,>=stealth]

\usetikzlibrary{intersections}
\usepackage{tkz-euclide}
\usetkzobj{all}
\usetikzlibrary{decorations.pathreplacing}

\tikzset{radiation/.style={{decorate,decoration={expanding waves,angle=90,segment length=4pt}}},
         relay/.pic={
        code={\tikzset{scale=5/10}
            \draw[semithick] (0,0) -- (1,4);% left line
            \draw[semithick] (3,0) -- (2,4);% right line
            \draw[semithick] (0,0) arc (180:0:1.5 and -0.5) node[above, midway]{#1};
            \node[inner sep=4pt] (circ) at (1.5,5.5) {};
            \draw[semithick] (1.5,5.5) circle(8pt);
            \draw[semithick] (1.5,5.5cm-8pt) -- (1.5,4);
            \draw[semithick] (1.5,4) ellipse (0.5 and 0.166);
            \draw[semithick,radiation,decoration={angle=45}] (1.5cm+8pt,5.5) -- +(0:2);
            \draw[semithick,radiation,decoration={angle=45}] (1.5cm-8pt,5.5) -- +(180:2);
  }}
}

\tikzset{radiation/.style={{decorate,decoration={expanding waves,angle=90,segment length=4pt}}},
         relay1/.pic={
        code={\tikzset{scale=7/10}

    \draw[thin] (0,0) -- (0,10);% left line
    \draw[thin] (5,0) -- (5,10);% right line
    \draw[thin] (0,0) -- (5,0);% left line
    \draw[thin] (0,10) -- (5,10);% right line

    \draw[thin] (0.75,1.5) -- (0.75,8.5);% left line
    \draw[thin] (0.75,8.5) -- (4.25,8.5);% right line
   \draw[thin] (4.25,8.5) -- (4.25,1.5);% left line
   \draw[thin] (4.25,1.5) -- (0.75,1.5);% right line

   \draw[thin] (2.5,0.7) circle(15pt);
   \draw[thin] (2.5,9.5) circle(10pt);

       \draw[thin] (1.5,9) -- (3.5,9);% right line
       \draw[thin] (1.5,8.7) -- (3.5,8.7);
       \draw[thin] (1.5,9) -- (1.5,8.7);
       \draw[thin] (3.5,9) -- (3.5,8.7);

  }}
}

\tikzset{
  block/.style={draw,text width=2em,minimum height=1em,align=center},
  arrow/.style={->}
}

\usepackage{algorithm,algorithmic}

\usetikzlibrary{arrows.meta,calc,decorations.markings,math,arrows.meta}
\usetikzlibrary{shapes.multipart}

\usepackage{tikz}
\usetikzlibrary{calc}
\usepackage{zref-savepos}

\usepackage[T1]{fontenc}% optional T1 font encoding

\ifCLASSINFOpdf
\else
\fi

\usepackage{amsmath}

\usepackage[cmintegrals]{newtxmath}

\begin{document}
\title{A Probabilistic Approach to Model SIC based RACH mechanism for Massive Machine Type Communications in Cellular Networks}
\author{Yeduri Sreenivasa Reddy,~\IEEEmembership{Student Member,~IEEE,}
        Ankit Dubey,~\IEEEmembership{Member,~IEEE,}
       Abhinav Kumar,~\IEEEmembership{Member,~IEEE,}
       and Trilochan Panigrahi,~\IEEEmembership{Member,~IEEE}
       
\thanks{This work was supported in part by the Science and Engineering Research Board (SERB), Govt. of India through its Early Career Research (ECR) Award (Ref. No. ECR/2016/001377), and the Department of Science and Technology (DST), Govt. of India (Ref. No. TMD/CERI/BEE/2016/059(G)).}       
\thanks{Y. S. Reddy, A. Dubey, and T. Panigrahi are with the Department of Electronics and Communication Engineering,
National Institute of Technology-Goa, Ponda 403401, India (e-mail:
ysreenivasareddy@nitgoa.ac.in;
ankit.dubey@nitgoa.ac.in;
tpanigrahi@nitgoa.ac.in).}
\thanks{A. Kumar is with the Department of Electrical Engineering,
IIT Hyderabad, Hyderabad 502285, India (e-mail: 
abhinavkumar@iith.ac.in).}}
%\markboth{IEEE INTERNET OF THINGS JOURNAL,~Vol.~14, No.~8, August~2015}%
%{Shell \MakeLowercase{\textit{et al.}}: Bare Demo of IEEEtran.cls for IEEE Communications Society Journals}
\maketitle
\begin{abstract}
In a cellular Internet of Things, burst transmissions from millions of machine type communications (MTC) devices can result in channel congestion. 
The main bottleneck in such
scenario is inefficient random access channel (RACH) mechanism that is used to attach MTC devices to a base station (BS). 
To address this issue of congestion in RACH mechanism, 3GPP has proposed an extended access barring (3GPP-EAB) mechanism. However, several works indicate that the performance of the 3GPP-EAB mechanism can be further improved. In this work, a successive interference cancellation (SIC) based RACH mechanism is considered to significantly increase the success rate and reduce congestion. In the proposed mechanism, the devices are allowed to transmit repeatedly for a finite number of times in a given cycle, and thereafter, the success rate is improved by applying back-and-forth SIC at the BS. A novel probabilistic approach of the proposed mechanism is presented with all transition and steady-state probabilities. Further, the probability of SIC for a given slot is derived. Through extensive numerical results, it is shown that the proposed mechanism significantly outperforms the existing ones in terms of the success rate. Moreover, to obtain the maximum success rate, the optimum number of devices to be entered in a cycle is also calculated.
\end{abstract}
\IEEEpeerreviewmaketitle
\section{introduction}
In a cellular Internet of Things (C-IoT), millions of MTC devices can connect to a single base station (BS). Typically, the MTC devices are small in size and can even be deployed in remote locations to collect data. 
These devices need to be energy efficient as they may have to run for years with limited battery.
The MTC devices wake up to transmit a small amount of data and go back to sleep state once the data is transmitted~\cite{atzori},~\cite{dama7}. For this data transmission to happen successfully, the device needs to attach to a BS for any information exchange.
This attachment mechanism, where the MTC device moves from radio resource control (RRC)-idle to RRC-connected state, is called random access channel (RACH) mechanism~\cite{Bezerra}.

The RACH mechanism involves four message exchanges between any MTC device and the BS~\cite{3gpp3}.
Prior to this mechanism, the BS broadcasts a message called the System Information Block (SIB)-$2$ that indicates the preamble group from which the contending MTC devices have to select preambles for transmission. 
Each contending MTC device randomly activates one of the $K$ preambles from the assigned group and transmits the same to the BS as \textit{msg}-$1$ in the uplink (UL).
Upon receiving \textit{msg}-$1$, the BS can only detect the number of active preambles but not the devices that have activated these preambles~\cite{Pratas1}.
To each active preamble, the BS responds with UL grant message (Random Access Response) denoted as \textit{msg}-$2$ in Physical Downlink Shared Channel (PDSCH) during response window. The \textit{msg}-$2$ indicates the resources assigned for transmitting \textit{msg}-$3$.
In case, an MTC device does not receive \textit{msg}-$2$ within some stipulated time, it restarts the mechanism after a random backoff time. Post $m$ such unsuccessful transmissions, the MTC device drops the data packet and restarts the RACH mechanism after a random backoff time.
Given, the MTC device has successfully processed \textit{msg}-$2$, it transmits \textit{msg}-$3$ using Physical Uplink Shared Channel (PUSCH). 
In case only one device activates a given preamble in a given radio frame (during \textit{msg}-$1$ transmission), the BS responds with an RRC-connected message (contention resolution) denoted by \textit{msg}-$4$ using Physical Downlink Control Channel (PDCCH). 
However, if more than one device activates the same preamble, then it leads to a collision in \textit{msg}-$3$. Then, such devices fail to receive \textit{msg}-$4$. Further, if an MTC device does not receive \textit{msg}-$4$ within some stipulated time, it restarts the RACH mechanism after a random backoff time~\cite{Bezerra},~\cite{mad3}. 

The existing 3GPP RACH mechanism is inefficient in presence of a large number of MTC devices~\cite{Bezerra}.
In order to address this problem, for a given finite number of preambles, 3GPP has proposed an extended access barring (3GPP-EAB) mechanism that selectively controls the network congestion using access control barring (ACB) parameter transmitted over SIB-2~\cite{3gpp3}. To further improve the network throughput, a fast RACH mechanism (FRM) has been proposed in~\cite{dama7} that allows an optimized number of device accesses in a given radio frame to maximize the RACH success rate.

Several surveys on MTC suggest that a successive interference cancellation (SIC) based RACH mechanism can further improve the success rate~\cite{msali},~\cite{biral}. Following these surveys, a simulation supported analysis of SIC based RACH mechanism has been presented in~\cite{Reddy} sans any analytical expressions and models. In order to provide better insights, we propose a probabilistic model for SIC based RACH mechanism with analytical expressions for all transition and steady-state probabilities. The key contributions of the proposed work are as follows:
\begin{itemize}
\item The proposed RACH mechanism is represented using a Markov chain model and analytical expressions for all the corresponding state transition probabilities are derived.
\item The analytical expression of the probability of SIC of a given slot is derived.
\item The steady-state analysis of the Markov chain is performed and the steady-state probabilities are derived.
\item Separate numerical analyses are presented to compute, (1) the optimal number of devices to be entered in a cycle and (2) the optimal repetition rate for a device, in order to maximize the success rate.
\item Finally, a numerical analysis is presented to steady the effect of physical  layer impairments on the performance of the proposed mechanism.
\end{itemize}

The rest of the paper is outlined as follows. The related work is presented in Section~\ref{RW}.
The existing RACH mechanisms and the proposed RACH mechanism are discussed in Section~\ref{SM}.
The Markov model of the proposed RACH mechanism with all transition probabilities is presented in Section~\ref{ana}.
In Section~\ref{STD}, the steady-state probabilities are derived.
The performance of the proposed RACH mechanism is evaluated through extensive
simulations and analysis in Section~\ref{RES}. Finally, Section~\ref{CON} provides some concluding remarks along with the future scope.
\section{related work}\label{RW}

The exponential growth in the number of devices degrades the performance of LTE RACH mechanism in terms of energy consumption and access delay~\cite{laya15}. The main bottleneck in such scenario is inefficient RACH mechanism that is used to attach MTC devices to a BS. 
To address the issue of congestion in the RACH mechanism, 
the 3GPP-EAB mechanism has been proposed in~\cite{3gpp5} that controls the burst MTC traffic using ACB parameter transmitted over SIB-2. The analytical model of the 3GPP-EAB mechanism has been presented in~\cite{cheng6}. 
Numerous works claim that the performance of the 3GPP-EAB can be further improved. 
A survey on two level classification of access management techniques focusing on delay, energy, and the success rate has been presented in~\cite{Islam1}.
A joint optimization technique has been proposed in \cite{Harwahyu} to configure the parameters such as number of preamble transmissions, size of backoff windows, and number of subcarriers in each coverage enhancement level  in order to maximize the success probability under a target delay constraint.
An FRM mechanism has been proposed in~\cite{dama7} that adjusts the number of MTC RACH accesses per radio frame to maximize the success rate.
An adaptive ACB scheme that dynamically adjusts the ACB parameter has been proposed in~\cite{tavana21} to control the congestion from burst MTC traffic.
A load balancing based dynamic access control scheme to control the radio access network (RAN) overload in the presence of high and low priority MTC devices has been proposed in~\cite{Ali}.

A dynamic  backoff scheme to maximize the RACH success rate has been proposed in \cite{lin18} that dynamically adjust the backoff window size by estimating the number of RACH attempts.
A mechanism to estimate the  number of preambles at the BS to maximize the success rate for an unknown number of devices and their access probabilities has been proposed in~\cite{Choi}. 
A prioritized random access mechanism with dynamic access barring
parameter for providing the quality of service (QoS) has been proposed in~\cite{lin17} that pre-allocates the resources by categorizing MTC devices into groups.
A Time Distributed Initial Access (TDIA) scheme has been proposed in~\cite{Balasubramaniam} to group the MTC devices based on their International Mobile Equipment Identity (IMEI) number  and assign each group with some preambles and start time. 
It has been shown in~\cite{Balasubramaniam} that this improves the probability of successful preamble transmissions.
A distributed queue based mechanism has been proposed in~\cite{Awuor} to place the MTC devices in a logical queue and allow them to access the BS based on their position in the queue.
In \cite{luis}, the number of contending devices in a random access slot is estimated by using maximum likelihood and Bayesian approach. The estimation is carried out with the help of joint probability density function of the number of successful and collided preambles.
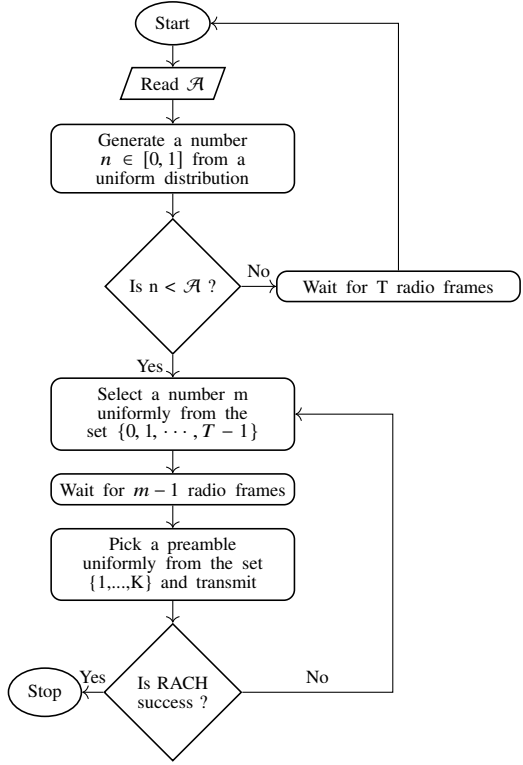
\begin{figure}
\centering
\begin{tikzpicture}[node distance=2cm]
\linespread{0.6}
\node (start) [startstop] {\scriptsize{Start}};
\node (in1) [io, below of=start, yshift = 1.2cm] {\scriptsize{Read $\cal{A}$}};

%\node (pro1) [process, below of=in1, yshift = 1.0cm,align = center] {\scriptsize{Generate a number $n \in [0,1]$} \\ \scriptsize{from a uniform distribution} };

\node (pro1) [process, below of=in1, yshift = 1.0cm,align = center,text width=3cm] {\scriptsize{Generate a number $n \in [0,1]$ from a uniform distribution}};

%\node (pro1) [process, below of=in1, yshift = 1.1cm,align = center] {\scriptsize{Generate n uniformly} \\ \scriptsize{distributed in [0,1]} };
\node (dec1) [decision, below of=pro1,align=center , yshift = 0.3cm] {\scriptsize{Is n \textless~$\cal{A}$ ?}};
\node (pro2a) [process, below of=dec1,yshift = 0.3cm, align = center,text width=3cm] {\scriptsize{Select a number m uniformly from the set $\{0, 1, \cdots, T-1\}$}};
\node (pro2b) [process, right of=dec1, xshift=1cm, align=center,text width=3cm] {\scriptsize{Wait for T radio frames}};
\node (pro3) [process, below of=pro2a, yshift = 1cm,text width = 3cm, align=center] {\scriptsize{Wait for $m-1$ radio frames}};
\node (pro4) [process, below of=pro3, yshift = 1cm, align = center,text width=3cm]  {\scriptsize{Pick a preamble uniformly from the set \{1,...,K\} and transmit}};
\node (dec2) [decision, below of=pro4,yshift=0.3cm, align=center] {\scriptsize{Is RACH} \\ \scriptsize{success ?}};
%\node (pro5) [process, below of=dec2, yshift=-0.9cm] {apply SIC  at other locations};
%\node (example-align1) [draw, align=left,below of=pro4]{example \\ example example};
\node (stop) [startstop, left of=dec2,xshift=0.3cm] {\scriptsize{Stop}};
%\node (pro4b) [process, right of=dec2, xshift=2.5cm] {Wait for T Radio frames};
\draw [arrow] (start) -- (in1);
\draw [arrow] (in1) -- (pro1);
\draw [arrow] (pro1) -- (dec1);
\draw [arrow] (dec1) -- node[left] {\scriptsize{Yes}}  (pro2a);
\draw [arrow] (dec1) -- node[above] {\scriptsize{No}} (pro2b);
\draw [arrow] (pro2a) -- (pro3);
\draw [arrow] (pro3) -- (pro4);
\draw [arrow] (pro4) -- (dec2);
\draw [arrow] (dec2) -- node[above] {\scriptsize{Yes}} (stop);
%\draw [arrow] (dec2) -| node {No} (pro2a);
%\draw[Line] ([yshift=5pt] dec2.east) -| (pro2b);
\draw [arrow] (pro2b) |- (start);
%\draw [arrow] (pro4) -- (stop);
\draw [arrow]
    (dec2.east) -- node[above] {\scriptsize{No}}
    ++(2.0cm,0) |-
    (pro2a.east)
    ;
%\draw [ultra thick, orange, ->] (m6.west) -- ++(-2.0cm,0);

\end{tikzpicture}
\caption{The 3GPP Extended Access Barring mechanism~\cite{dama7}.}
\label{Fig. 3GPP}
\end{figure}

A random access scheme for generating virtual preambles
by adding Physical-RACH indexes to the physical preambles
has been proposed in~\cite{kim16}
to increase the number of available
preambles.
An energy efficient random access scheme has been proposed in~\cite{Beyene} that allows the devices to send small packets in order to reduce the signal overhead and to create more resources for data transmissions.
An efficient cell planning and Zadoff-Chu root sequence allocation technique has been proposed in~\cite{Kalalas} that efficiently allocates the root sequences among multiple cells to reduce the inter-cell interference.
An improved group-paging mechanism has been proposed in~\cite{Arouk} to scatter the paging operation over  a group paging interval, instead of letting all the MTC devices to contend at the same time to improve the channel access probability.
An analytical model to analyse the performance of RACH with MTC traffic following a beta distribution has been proposed in~\cite{arouk19}. 
A simulation supported analysis of SIC based multiple access schemes for MTC have been presented in \cite{Reddy}, \cite{ye}, \cite{Interdonato}.
However, to the best of our knowledge no work exists that presents the analysis of improved RACH mechanism with SIC.
\begin{figure}
\begin{center}
\begin{tikzpicture}[node distance=2cm]
\linespread{0.6}
\node (start) [startstop] {\scriptsize{Start}};
\node (in1) [io, below of=start, yshift = 1.2cm] {\scriptsize{Read $\cal{B}$, W, X}};
\node (pro1) [process, below of=in1, yshift = 1cm,align = center,text width=3cm] {\scriptsize{Generate a number $n \in [0,1]$ from a uniform distribution} };
\node (dec3) [decision, right of=pro1, xshift=1.0cm, align = center] {\scriptsize{Is $\cal{B}$\! \textgreater $\frac{1}{W}$?}};
\node (dec1) [decision, below of=pro1, yshift=0.15cm, align = center] {\scriptsize{Is n \textless ~$\cal{B}$ W ?}};
\node (pro2a) [process, below of=dec1, yshift=0.15cm,align = center,text width=3cm]  {\scriptsize{Pick a preamble uniformly from the set \{1,...,K\} transmit}};
%\node (pro3) [process, below of=pro2a] {Generate d using a distribution};
\node (pro2b) [process, right of=dec1, xshift=1.0cm] {\scriptsize{Update $\cal{B}$}};
%\node (pro4) [process, below of=pro3] {Pick d radio frames out of [0,T-1]};
\node (dec2) [decision, below of=pro2a,yshift=0.3cm, align = center] {\scriptsize{Is RACH} \\ \scriptsize{success ?}};
%\node (pro5) [process, below of=dec2, yshift=-0.9cm] {apply SIC  at other locations};
%\node (example-align1) [draw, align=left,below of=pro4]{example \\ example example};
\node (stop) [startstop, left of=dec2, xshift=0.1cm] {\scriptsize{Stop}};
%\node (pro4b) [process, right of=dec2, xshift=2.5cm] {Wait for T Radio frames};
\draw [arrow] (start) -- (in1);
\draw [arrow] (in1) -- (pro1);
\draw [arrow] (pro1) -- (dec1);
\draw [arrow] (dec1) -- node[left] {\scriptsize{Yes}}  (pro2a);
\draw [arrow] (dec1) -- node[above] {\scriptsize{No}} (pro2b);
\draw [arrow] (pro2a) -- (dec2);
\draw [arrow] (pro2b) -- (dec3);
\draw [arrow] (dec3) -- node[above] {\scriptsize{No}} (pro1);
\draw [arrow] (dec2) -- node[above] {\scriptsize{Yes}} (stop);
\draw [arrow] (dec2) -| node[right] {\scriptsize{No}} (pro2b);
\draw [arrow] (dec3) |- node[right] {\scriptsize{Yes}} (in1);
%\draw [arrow] (dec2) -- (stop);
\end{tikzpicture}
\caption{The Fast RACH mechanism~\cite{dama7}.}
\label{FRM}
\end{center}
\end{figure}
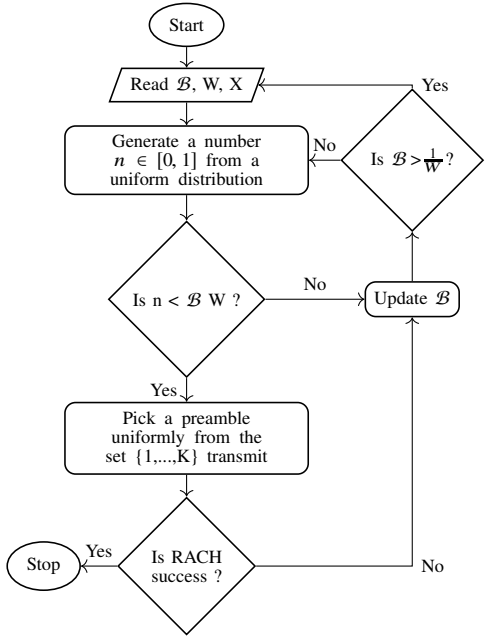

The novelty of the proposed mechanism in comparison with the existing ones lies in the utilization of SIC.
Next, we discuss various RACH mechanisms that are considered in this work.
\section{rach mechanisms} \label{SM}
In this section, we discuss the proposed SIC based RACH mechanism along with the existing 3GPP-EAB and FRM mechanisms. 
\subsection{3GPP-EAB Mechanism}
The 3GPP-EAB, as given in~\cite{dama7},~\cite{3gpp5}, is depicted with a flowchart in Fig.~\ref{Fig. 3GPP}. In 3GPP-EAB, initially, the BS broadcasts an ACB parameter $\cal{A}$\,$\in$\,$[0,1]$ that significantly controls the number of RACH accesses. After successful reception of $\cal{A}$, each MTC device generates a number $n$\,$\in$\,$[0,1]$ from a uniform distribution. 
All the MTC devices with $n$\,$<$\,$\cal{A}$ enter the contention loop of $T$ slots by selecting a number $m$ uniformly from the set $\{0, 1, \cdots, T-1\}$.
All the other MTC devices with $n$\,$>$\,$\cal{A}$ wait for $T$ slots and repeat the mechanism. A device with number $m$ waits for $m-1$ slots and transmits 
a preamble that is uniformly selected from the set $\{1, 2, \cdots, K\}$
as \textit{msg}-$1$ in slot $m$. In case the device successfully completes the transmission in slot $m$ then it is notified via \textit{msg}-$4$ by the BS. Otherwise, it selects a new $m$ uniformly from the set $\{0,1,\cdots,T-1\}$ and waits for next $m-1$ slots and transmits a fresh preamble that is uniformly selected from the set
$\{1, 2, \cdots, K\}$
in slot $m$. This mechanism is repeated until all the devices get access to BS.
\subsection{Fast RACH Mechanism}
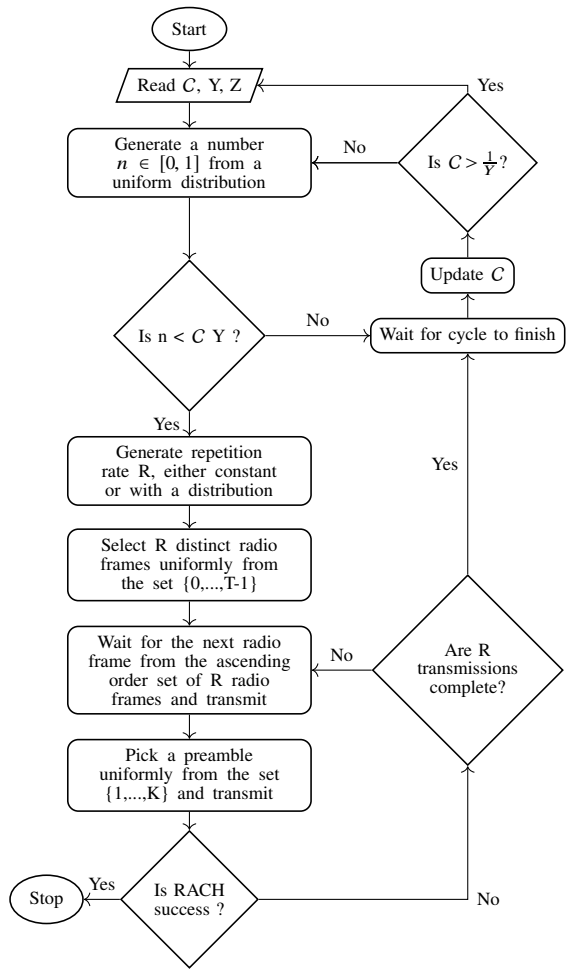
\begin{figure}
\begin{center}
\begin{tikzpicture}[node distance=2cm]
\linespread{0.6}
\node (start) [startstop] {\scriptsize{Start}};
\node (in1) [io, below of=start, yshift = 1.25cm] {\scriptsize{Read ${\cal{C}}$, Y, Z}};
\node (pro1) [process, below of=in1, yshift = 0.97cm, align = center,text width=3cm] {\scriptsize{Generate a number $n \in [0,1]$ from a uniform distribution} };
\node (dec1) [decision, below of=pro1,yshift=-0.3cm, align = center] {\scriptsize{Is  n \textless  ~${\cal{C}}$ Y ?}};
\node (pro2a) [process, below of=dec1, yshift=0.2cm,align = center,text width=3cm] {\scriptsize{Generate repetition rate R, either constant or with a  distribution}};

%{\scriptsize{Pick a preamble uniform randomly} \\ \scriptsize{from \{1,...,K\}}};

%\node (pro6) [process, below of=dec6] {Update P};
\node (pro6) [decision, right of=pro1, xshift = 1.7cm] {\scriptsize{Is ${\cal{C}}\!>\!\frac{1}{Y}$? }};
%\node (pro6) [process, below of=dec6, yshift = -0.25cm] {Update P};
\node (pro3) [process, below of=pro2a,align = center, yshift = 0.75cm,text width=3cm] {\scriptsize{Select R distinct radio frames uniformly from the set \{0,...,T-1\}}};
\node (pro2b) [process, right of=dec1, xshift=1.7cm] {\scriptsize{Wait for cycle to finish}};
\node (update) [process, above of=pro2b, yshift=-1.2cm] {\scriptsize{Update ${\cal{C}}$}};
%\node (pro4) [process, below of=pro3,align = center, yshift = 0.8cm] {\scriptsize{Wait for the next radio frame}\\ \scriptsize{from the ascending order set of} \\ \scriptsize{R radio frames and transmit}};

\node (pro4) [process, below of=pro3,align = center, yshift = 0.6cm,text width=3cm] {\scriptsize{Wait for the next radio frame from the ascending order set of R radio frames and transmit}};

\node (pro120) [process, below of=pro4,align = center, yshift = 0.6cm,text width=3cm] {\scriptsize{Pick a preamble uniformly from the set \{1,...,K\} and transmit}};
\node (dec2) [decision, below of=pro120, align = center, yshift = 0.35cm] {\scriptsize{Is RACH} \\ \scriptsize{success ?}};
\node (dec3) [decision, right of=pro4, align = center, xshift = 1.7cm] {\scriptsize{Are R}\\
\scriptsize{transmissions} \\ \scriptsize{complete?}};
%\node (pro5) [process, below of=dec2, yshift=-0.9cm] {apply SIC  at remaining d-1 locations};
%\node (example-align1) [draw, align=left,below of=pro4]{example \\ example example};
\node (stop) [startstop, left of=dec2, xshift= 0.1cm] {\scriptsize{Stop}};
%\node (pro4b) [process, right of=dec2, xshift=2.5cm] {Wait for T Radio frames};
\draw [arrow] (start) -- (in1);
\draw [arrow] (in1) -- (pro1);
\draw [arrow] (pro1) -- (dec1);
\draw [arrow] (dec1) -- node[left] {\scriptsize{Yes}}  (pro2a);
\draw [arrow] (dec1) -- node[above] {\scriptsize{No}} (pro2b);
\draw [arrow] (pro6) -- node[above] {\scriptsize{No}} (pro1);
\draw [arrow] (pro6) |- node[right] {\scriptsize{Yes}}  (in1);
\draw [arrow] (pro2a) -- (pro3);
\draw [arrow] (pro3) -- (pro4);
\draw [arrow] (pro4) -- (pro120);
\draw [arrow] (pro120) -- (dec2);
\draw [arrow] (dec2) -- node[above] {\scriptsize{Yes}} (stop);
\draw [arrow] (dec3) -- node[left] {\scriptsize{Yes}} (pro2b);
\draw [arrow] (dec3) -- node[above] {\scriptsize{No}} (pro4);
%\draw [arrow] (dec2) -| node {No} (pro2b);
\draw [arrow] (dec2) -| node[right] {\scriptsize{No}} (dec3);
\draw [arrow] (pro2b) -- (update);
\draw [arrow] (update) -- (pro6);
\draw [arrow] (pro6) -- (pro1);
%\draw [arrow] (dec6) -- node[above] {No} (pro1);
%\draw [arrow] (dec6) |- node[above] {yes} (in1);
%\draw [arrow] (pro5) -- (stop);
\end{tikzpicture}
\caption{The Proposed SIC based RACH mechanism~\cite{Reddy}.}
\label{Prop}
\end{center}
\end{figure}
In the Fast RACH mechanism, as given in~\cite{dama7}, the BS broadcasts a number $\cal{B}$\,$\in$\,$[0,1]$ that is the reciprocal of the average number of active devices as shown in the flowchart in Fig.~\ref{FRM}. Post reading $\cal{B}$, 
all MTC devices also read numbers $W$ and $X$ which are transmitted along with $\cal{B}$.
These parameters, $W$ and $X$, are added to control the number of contending devices in each slot to improve the success rate. The values of $W$ and $X$ represent the average number of RACH accesses
and corresponding average RACH successes in a given radio frame, respectively.
These values can also be inbuilt in the MTC device itself. Next, each MTC device generates a number $n$\,$\in$\,$[0,1]$ from a uniform distribution. All the devices with $n$\,$<$\,${\cal{B}}\,W$ enter into the radio frame by uniformly selecting a preamble from the set $\{1, 2, \cdots, K\}$ and transmit the same. 
In case the transmission goes successful, the device is notified via \textit{msg}-$4$. 
With ${\cal{B}}_{t}$ being the present value of $\cal{B}$, all the unsuccessful MTC devices 
repeat the RACH mechanism with ${\cal{B}}_{t+1}$
which is the updated value of $\cal{B}$ and is given as
\begin{equation*} \label{a_3}
{\cal{B}}_{t+1} = \min\left\{\frac{1}{W},\frac{1}{\max\left\{1,\frac{1}{{\cal{B}}_t}-X\right\}}\right\}\,, 
\end{equation*}
where, ${\cal{B}}_{t+1}$ is the reciprocal of the average number of unsuccessful devices. The number of unsuccessful devices is obtained by subtracting the average number of RACH successes in a radio frame ($X$) from the number of devices contending in the previous radio frame, i.e., $1/{\cal{B}}_t$.
Alternatively, whenever ${\cal{B}}_{t+1}\!>\! 1/X$, all the remaining MTC devices read
$\cal{B}$ directly from the BS.
This is to ensure that all the devices are allowed to contend whenever the number of
devices is less than the number of available preambles.
\subsection{Proposed SIC Based RACH Mechanism}\label{proposed_algo}
In the proposed mechanism, the BS broadcasts ${\cal{C}}\in [0,1]$ over SIB-$2$ to all the MTC devices 
in its coverage, where, ${\cal{C}}$
is the reciprocal of the average number of active devices.
The BS also broadcasts numbers
$Y$ and $Z$ along with $D$, where, $Y = E\,K\,T$.
Here, the parameters, $E$, $K$, and $T$, are added to control the number of contending devices in each cycle to improve the success rate. 
The values of $K$ and $T$ represent the number of available preambles and the number of slots in each cycle, respectively, same as in the 3GPP-EAB mechanism. Further, $E$ is the fractional value that controls the number of contending devices in each cycle. The parameters $Y$ and $Z$ represent the average number of RACH accesses and corresponding RACH
successes in a given cycle, respectively, in the proposed mechanism for a given value of $E$.
After successful reception of SIB-$2$, each MTC device generates a number $n$\,$\in$\,$[0,1]$ from a uniform distribution. 
All the MTC devices with $n$\,$<$\,${\cal{C}}\,Y$ enter into the contention loop of $T$ slots. All the other MTC devices with $n$\,$>$\,${\cal{C}}\,Y$ wait for $T$ slots and restart the contention mechanism.
A device that enters the contention loop uniformly selects $R$ ($R<T$) distinct numbers from the set $\{0, 1, 2, \cdots, T-1\}$ and creates an ordered set $\{m_1, m_2, \cdots, m_R\}$ where, $m_1<m_2<\cdots<m_R$. Here, $R$ represents the repetition rate of a device. 
A device with the first number as $m_1$ waits for $m_1$ slots and transmits a preamble 
that is uniformly selected from the set $\{1, 2, \cdots, K\}$ 
as \textit{msg}-$1$  in slot $m_1$. The successful transmission of a device is notified via \textit{msg}-$4$. In case the device is successful then it terminates its further transmissions. Otherwise, it waits for $(m_2-m_1-1)$
slots and transmits a 
fresh preamble that is uniformly selected from the  set $\{1, 2, \cdots, K\}$ in slot $m_2$. 
With ${\cal{C}}_{c}$ being the present value of ${\cal{C}}$ in the current cycle,
the remaining MTC devices that have not been successful in this cycle repeat the RACH mechanism with ${\cal{C}}_{c+1}$ which is an updated value of ${\cal{C}}$ and is given as
 \begin{equation*}
{\cal{C}}_{c+1} = \min\left\{\frac{1}{Y},\frac{1}{\max\left\{1,\frac{1}{{\cal{C}}_{c}}-Z\right\}}\right\}\,, \label{A_u}
\end{equation*}
where, ${\cal{C}}_{c+1}$ is the reciprocal of the average number of remaining unsuccessful MTC devices. The 
count of all the remaining MTC devices
is obtained by subtracting the average number of RACH successes in a cycle ($Z$) from the number of devices contending in the previous cycle, i.e., $1/{\cal{C}}_c$.
This process continues until all the devices get access to the BS.
Alternatively, whenever ${\cal{C}}_{c+1}$\,$>$\,$ 1/Y$, all the remaining MTC devices read
${\cal{C}}$ directly from the BS. This is to ensure that all the devices are allowed to contend whenever the number of
devices is less than the number of available preambles.
\textit{Note
that, throughout the text, we use the terms radio frame and slot interchangeably.}

Upon successful reception of \textit{msg}-$3$ from a device, the BS applies SIC at all the previously transmitted slots of that particular device.
The BS can retrieve the \textit{msg}-$3$ of another device, if this device alone shares a preamble with the above successful device in any of the previously transmitted slots as shown in~\cite{Wei16}.
Given $r$ devices are successful with RACH, the \textit{msg}-$3$ of a device that has concurrently shared the preamble with these $r$ devices can be successfully obtained at the BS with SIC.
Here, \textit{msg}-$3$ indicates the device identity which is unique for a particular device as given in~\cite{Cherk16}.
The Markov chain model of the proposed SIC based RACH mechanism with analytical expressions of all the corresponding transition probabilities is discussed in the next section.
\def\radius{3.5mm}
\begin{figure*}
	\centering
	
%\resizebox{13cm}{10cm}
%{

	\begin{tikzpicture}[>=latex]
	\tikzset{node style/.style={state, ellipse,semithick}}
	
	%   \tikzstyle{stateinv}=[inner sep=5pt]
	\tikzstyle{point}=[coordinate]
	\tikzstyle{stateEdgePortion} = [black,semithick];
	\tikzstyle{stateEdge} = [stateEdgePortion,->,semithick];
	\tikzstyle{edgeLabel} = [pos=0.5, text centered, font={\sffamily\small}];
	
%	\node[node style]               (CCA)   {0,-1};
	\node[node style]   (b00)  {0,0};
	
	\node[point, above=of b00,yshift=0.3cm]  (b0--1) {};
%	\node[node style, above=of b0--1]  (b0-1) {-1,0};
	
	\node[node style, above=of b0--1 , label=center:{-1,0}] (b0-1) {\phantom{0,0}};
	%\node[node style, above=of b00]  (b0-1) {-1,-1};
	\node[node style, right=of b00,xshift=0.3cm]  (b01) {0,1};
	\node[point, left=of b00]  (pointnode) {};
	\node[point, left=of b0-1,xshift=-1.40cm]  (pointnode_1) {};
	\node[node style, right=of b01,xshift=0.3cm]  (b02) {0,2};
	\node[draw=none,  right=of b02,xshift=0.3cm]   (b02-b0wp) {$\cdots$};
%	\node[node style, right=of b02-b0wp,xshift=0.3cm] (b0wp)   {0,T-2};

	\node[node style, right=of b02-b0wp,xshift=0.3cm , label=center:{0,T-2}] (b0wp) {\phantom{0,0}};

%	\node[node style, right=of b0wp,xshift=0.3cm] (b0w)   {0,T-1};
	
		\node[node style, right=of b0wp,xshift=0.3cm , label=center:{0,T-1}] (b0w) {\phantom{0,0}};	
	%\node[point, above=of b00]   (a)    {};
%	\coordinate[draw=none,left=of CCA]          (2CCA);

%	\node[point, below=of CCA]   (11)   {};
	\node[point, below=of b00]  (11)  {};
	\node[point, below=of b01] (12) {};
	\node[point, below=of b02] (13) {};
%	\node[point, below=of b02-b0wp] (12) {};
	\node[point, below=of b0wp]   (16)   {};
	\node[point, below=of b0w]   (17)    {};

%	\node[node style, below=of 11]   (CCA1)  {1,-1};
	\node[node style, below=of 11,yshift=-0.3cm]  (b10) {1,0};
	\node[node style, left=of b10]  (succ) {4,0};
	\node[node style, below=of 12,yshift=-0.3cm]  (b11)   {1,1};
	\node[node style, below=of 13,yshift=-0.3cm]  (b12)   {1,2};
	\node[draw=none,  right=of b12,xshift=0.3cm]   (b12-b1wp) {$\cdots$};
%	\node[node style, below=of 16,yshift=-0.3cm] (b1wp)   {1,T-2};
	
	\node[node style, below=of 16,yshift=-0.3cm , label=center:{1,T-2}] (b1wp) {\phantom{0,0}};	
	%\node[node style, below=of 17] (b1w)   {$1,T-1$};
	%\node[node style, below=of succ]  (succ_1) {4,0};

%	\node[point, below=of CCA]   (11)   {};
	\node[point, below=of b10]  (21)  {};
	\node[point, below=of b11] (22) {};
	\node[point, below=of b12] (23) {};
%	\node[point, below=of b02-b0wp] (12) {};
	\node[point, below=of b1wp]   (26)   {};
%	\node[point, below=of b1w]   (27)    {};

%	\node[node style, below=of 11]   (CCA1)  {1,-1};
	\node[node style, below=of 21,yshift=-0.3cm]  (b20) {2,0};
	\node[node style, below=of 22,yshift=-0.3cm]  (b21)   {2,1};
	\node[node style, below=of 23,yshift=-0.3cm]  (b22)   {2,2};
	\node[draw=none,  right=of b22,xshift=0.3cm]   (b22-b2wp) {$\cdots$};
%	\node[node style, below=of 26,yshift=-0.3cm] (b2wp)   {2,T-2};
	
	\node[node style, below=of 26,yshift=-0.3cm, label=center:{2,T-2}] (b2wp) {\phantom{0,0}};	
	%\node[node style, below=of 27] (b2w)   {$2,T-1$};
	
	\node[point, below=of b20,yshift=-0.3cm]  (ccr1) {};
	\node[point, left=of b20]  (ccr12) {Wt};
	%\node[node style, left=of ccr12]  (ccr13) {Wt,0};
	\node[point, below=of b21,yshift=-0.3cm]  (ccr2) {};
	\node[point, below=of b22,yshift=-0.3cm]  (ccr3) {};
	\node[point, below=of b2wp,yshift=-0.3cm]  (ccr4) {};
	%\node[point, below=of b2w]  (ccr5) {};
	\node[point, below=of b2wp,yshift=-0.3cm]   (timeWait)    {}; 
	
	\node[node style, below=of ccr1,yshift=-0.3cm]  (b40) {3,0};
	\node[node style, right=of b40,xshift=0.3cm]  (b41)   {3,1};
	\node[node style, right=of b41,xshift=0.3cm]  (b42)   {3,2};
	\node[draw=none,  right=of b42,xshift=0.3cm]   (b42-b4wp) {$\cdots$};
%	\node[node style, right=of b42-b4wp,xshift=0.3cm] (b4w)   {3,T-3};	
	
	\node[node style, right=of b42-b4wp,xshift=0.3cm, label=center:{3,T-3}] (b4w) {\phantom{0,0}};		
	
	\node[point, left=of b40,xshift=-1.51cm] (fds) {};
	%\node[point, left=of fds,xshift=-0.3cm] (fds1) {};

		\coordinate (timeWait2ClosedA) at ($(timeWait.south) + (0,0)$);
	\coordinate (timeWait2ClosedB) at ($(timeWait.south -| b20.west) + (-1.59em,0)$);
%	\coordinate (timeWait2ClosedC) at ($(succ.north -| succ.west) + (-2.68em,-0.1em)$);
%	\coordinate (timeWait2ClosedD) at ($(succ.north) + (0,0)$);
	\coordinate (timeWait2Closedx) at ($(succ.north -| b0-1.west) + (-4.09em,16.3em)$);
	
	\draw[semithick] (succ.north) edge node[left] {1}(timeWait2Closedx);

%	(ccr12)  edge[bend left=40] node[left] {$1-P_{SIC}$}        (pointnode)

		\node[point, below=of b40,yshift=-0.3cm]  (b50) {};
	%\node[point, left=of b20]  (ccr12) {Wt};
	%\node[node style, left=of ccr12]  (ccr13) {Wt,0};
	\node[point, below=of b41,yshift=-0.3cm]  (b51) {};
	\node[point, below=of b42,yshift=-0.3cm]  (b52) {};
	%\node[point, below=of b4w]  (b5w) {};
\node[point, below=of b4w,yshift=-0.3cm]   (timeWait3)    {}; 	
\coordinate (timeWait2ClosedA1) at ($(timeWait3.south) + (0,0)$);
		\coordinate (timeWait2ClosedB1) at ($(timeWait3.south -| b20.west) + (-4.212em,0)$);
			\coordinate (timeWait2Closedx1) at ($(succ.north -| b0-1.west) + (-3.9em,13.40em)$);
%\draw (b20.west) edge (ccr12);	
	
\draw[semithick] (fds.west) edge node[left=0.95,anchor=center] {1-$P_{SIC}(3,0)$}        (pointnode_1);	

%\draw (b40.west) edge (fds);
%\draw (succ) |- edge (pointnode_1);	
	
%	
%		(ccr12)  edge[bend left=40] node[left] {$1-P_{SIC}$}        (pointnode)

%\draw (b20) edge node[right]{$P_C$}(succ);

	\draw[semithick] (timeWait3.south) edge (timeWait2ClosedA1);
\draw[semithick] (timeWait3.south) edge[stateEdgePortion] (timeWait2ClosedA1);
	\draw (timeWait2ClosedA1) edge 
	node[edgeLabel, text width=10.25em, yshift=0em,below]{}	
	(timeWait2ClosedB1);
	
\draw[semithick] (timeWait.south) edge (timeWait2ClosedA);
\draw[semithick] (timeWait.south) edge[stateEdgePortion] (timeWait2ClosedA);
%	\draw (timeWait2ClosedA) edge 
%	node[edgeLabel, text width=10.25em, yshift=0em,below]{$P_{SIC}(j)$}	
%	(timeWait2ClosedB);	

\draw[->,name path = line 1] (b20) -- (ccr1)node[above right  =0.7cm,anchor=center, yshift=0.15cm]{1-$P_{t2}$};	

\path[semithick,name path = line 2] (timeWait2ClosedA) -- (timeWait2ClosedB);	
 	  \path [semithick,name intersections={of = line 1 and line 2}];
  \coordinate (S)  at (intersection-1);
\path[semithick,name path=circle] (S) circle(\radius);

  \path [semithick,name intersections={of = circle and line 2}];
  \coordinate (I1)  at (intersection-1);
  \coordinate (I2)  at (intersection-2);
  
      \draw[semithick] (timeWait2ClosedA) edge node[edgeLabel, text width=10.25em, yshift=0em,below]{}  (I1);
  \draw[semithick] (I2)--  (timeWait2ClosedB)node[below right=0.02cm]{1};
  
  \tkzDrawArc[semithick,color=black](S,I1)(I2);

    \draw [->,semithick,name path = line 3] (timeWait2ClosedB1) -- (succ);
\path[semithick,name path = line 4] (b40) -- (fds);
   	  \path [semithick,name intersections={of = line 3 and line 4}];
  \coordinate (S)  at (intersection-1);
\path[semithick,name path=circle] (S) circle(\radius);
  \path [semithick,name intersections={of = circle and line 4}];
  \coordinate (I1)  at (intersection-1);
  \coordinate (I2)  at (intersection-2);
        \draw[semithick] (b40) edge node {}  (I1);
  \draw[semithick] (I2)--  (fds)node{};
 \tkzDrawArc[color=black,semithick](S,I1)(I2);

  \draw [->,semithick](timeWait2ClosedB) -- (succ);
  
  	\draw [->,semithick](timeWait2Closedx) -- (b0-1);

 \path[->]

	(b00)   edge node[right=0.3,anchor=center] {$P$}         (11)
	(11)   edge node[right =0.25,anchor=center,yshift=-0.06cm] {$\alpha_0$}                         (b10)
	(11)   edge node[right=0.35,anchor=center,xshift=0.06cm] {$\alpha_1$}        (b11)
	(11)   edge node[right=0.5,anchor=center,yshift=0.04cm,xshift=0.06cm] {$\alpha_2$}        (b12)
	(11)   edge node[right=1.05,anchor=center,yshift=0.02cm,xshift=0.2cm] {$\alpha_{T-2}$}        (b1wp)
	%(11)   edge[bend left=5] node[above] {$q_{T-1}$}        (b1w)
	
	(b0-1) edge[loop above] node[above=0.2,anchor=center] {$P_{Sleep}$} (b0-1)
	(b0-1)   edge node[right=1,anchor=center] {$1-P_{Sleep}$}         (b0--1)
	(b0--1)   edge node[right] {}         (b00)
	(b0--1)   edge node[right] {}         (b01)
	(b0--1)   edge node[right] {}         (b02)
	(b0--1)   edge node[right] {}         (b0wp)
	(b0--1)   edge[bend left =5,anchor=center] node[above] {$\frac{1}{T}$}         (b0w)
	(b10)   edge node[right=0.45,anchor=center] {1-$P_{t1}$}         (21)
	(21)   edge node[right=0.25,anchor=center,yshift=-0.06cm] {$\beta_0$}                         (b20)
	(21)   edge node[right=0.37,anchor=center,xshift=0.06cm] {$\beta_1$}                         (b21)
	(21)   edge node[right=0.45,anchor=center,yshift=0.04cm,xshift=0.08cm] {$\beta_2$}                         (b22)
	(21)   edge node[right=0.85,anchor=center,yshift=0.01cm,xshift=0.32cm] {$\beta_{T-2}$}        (b2wp)
	%(21)   edge[bend left=5] node[above] {$s_{T-1}$}        (b2w)
	
	(b00)  edge node[below=0.2,anchor=center] {1-$P$}        (b01)
	(b01)  edge node[below=0.2,anchor=center] {1}        (b02)
	(b02)  edge node[below] {} (b02-b0wp)
	(b02-b0wp) edge node[below] {}        (b0wp)
	(b0wp)  edge node[below=0.2,anchor=center] {1}        (b0w)
	(b0w)  edge[bend left=25,anchor=center] node[above] {1}        (b00)
	
	(b11)  edge node[below=0.2,anchor=center] {1}        (b10)
	(b12)  edge node[below=0.2,anchor=center] {1}        (b11)
	(b1wp)  edge node[below] {}        (b12-b1wp)
	(b12-b1wp) edge node[below=0.2,anchor=center] {}        (b12)
	(b10)  edge node[below=0.25,anchor=center] {$P_{t1}$}        (succ)

%	(ccr12)  edge node[right] {$P_{Wt}$}        (succ)

	(b21)  edge node[below=0.6,anchor=center] {1-$P_{SIC}(2,1)$}        (b20)
	(b22)  edge node[below=0.6,anchor=center] {1-$P_{SIC}(2,2)$}        (b21)
	(b2wp)  edge node[below] {}        (b22-b2wp)
	(b22-b2wp)  edge node[below] {}        (b22)
	
%	 (b20) edge node[right] {$1-P_{SIC}$}   (ccr1);
%	(b20)  edge node[right] {}        (ccr1)
	(b21)  edge node[right=0.8,anchor=center] {$P_{SIC}(2,1)$}        (ccr2)
	(b22)  edge node[right=0.8,anchor=center] {$P_{SIC}(2,2)$}        (ccr3)
	(b2wp)  edge node[right=1.2,anchor=center] {$P_{SIC}(2,T-2)$}        (ccr4)
	(pointnode_1) edge node[right] {}        (b0-1)

	(ccr1)   edge node[right=0.25,anchor=center,yshift=-0.055cm] {$\gamma_0$}                         (b40)
	(ccr1)   edge node[right=0.35,anchor=center] {$\gamma_1$}                         (b41)
	(ccr1)   edge node[right=0.5,anchor=center,yshift=0.02cm] {$\gamma_2$}                         (b42)
	(ccr1)   edge node[right=0.8,anchor=center,xshift=0.2cm] {$\gamma_{T-3}$}        (b4w)

	(b40)  edge node[right=0.8,anchor=center] {$P_{SIC}(3,0)$}        (b50)
		(b41)  edge node[right=0.8,anchor=center] {$P_{SIC}(3,1)$}        (b51)
	(b42)  edge node[right=0.8,anchor=center] {$P_{SIC}(3,2)$}        (b52)
	(b4w)  edge node[right=1.2,anchor=center] {$P_{SIC}(3,T-3)$}        (timeWait3)

		(b41)  edge node[below=0.6,anchor=center] {1-$P_{SIC}(3,1)$}        (b40)
	(b42)  edge node[below=0.6,anchor=center] {1-$P_{SIC}(3,2)$}        (b41)
	(b42-b4wp) edge node[below=0.2] {}        (b42)
	(b4w)  edge node[right] {}        (b42-b4wp)
%	(pointnode_1)  edge node[left] {1}        (b0-1)
	%(ccr12)  edge[bend left=40] node[left] {$1-P_{SIC}$}        (pointnode)

	(b20) edge node[right=0.4,anchor=center] {$P_{t2}$} (succ);

	%(pointnode) edge node[left] {}  (b00);

	\end{tikzpicture}
%}
	\caption{The Markov chain model for the Proposed SIC based RACH mechanism}
	\label{Mod12}
\end{figure*}
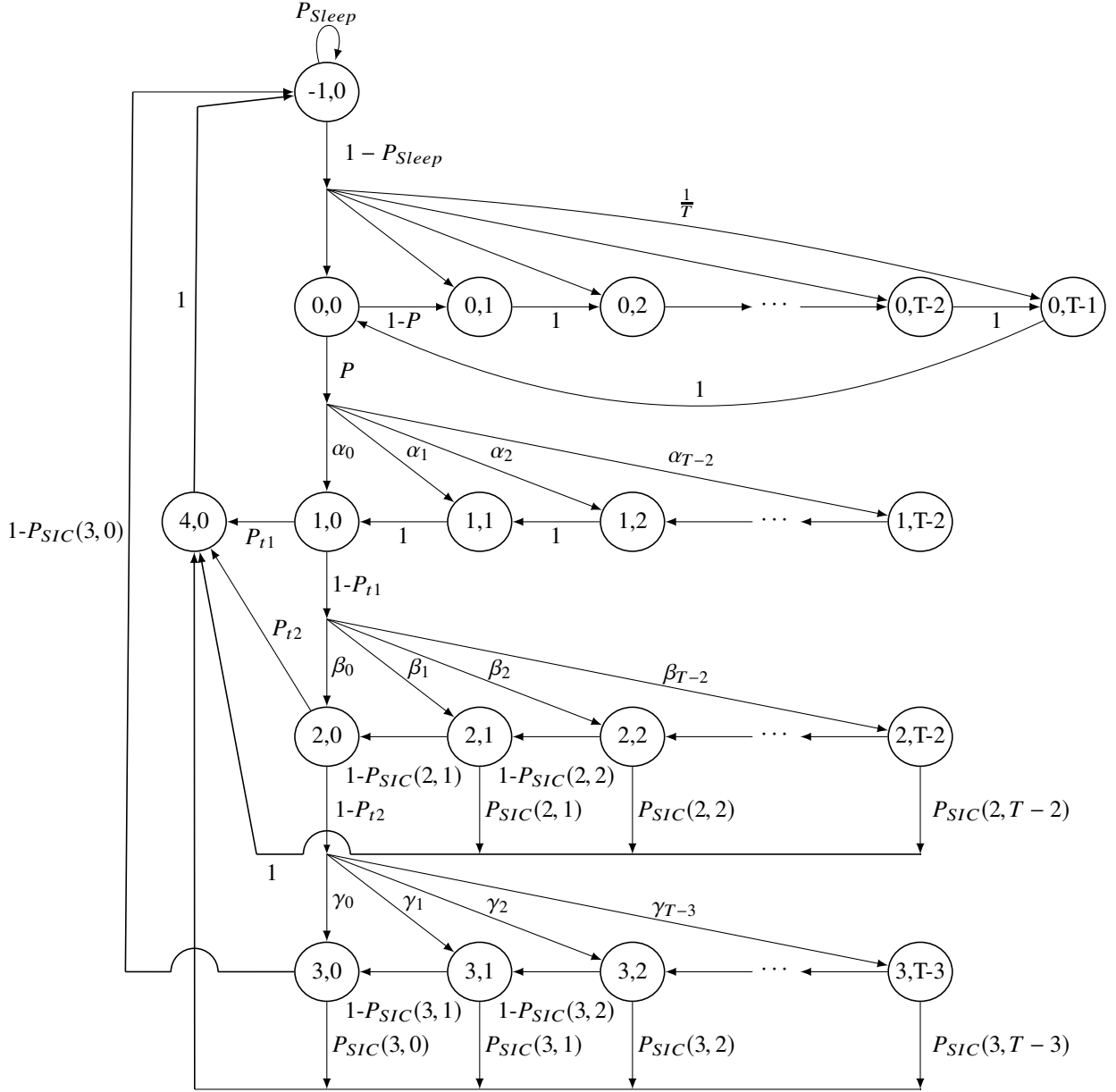
\section{analytical model} \label{ana}
The core contribution of this paper is the analytical evaluation
of the proposed SIC based RACH mechanism, under the assumption of ideal physical (PHY) channel conditions.
The analysis is divided into two parts. 
Initially, we study the behavior of a single MTC device with a Markov chain and then 
the steady-state analysis is carried out in the next section to derive the achievable success rate.  Let $s(t)\in \{-1,R+2\}$ and $b(t)\in \{0,T-1\}$ be the stochastic processes representing the backoff stage and backoff counter, respectively, for a given device as shown in Fig.~\ref{Mod12}. An integer time scale is adopted such that $t$ and $t+1$ correspond to the beginning of two consecutive time slots. Further, the backoff time counter of each device decrements at the beginning of each time slot. 
Moreover, a device passes through a maximum of $R+4$ stages for a given value of $R$.
First two stages, $s(t)=-1$ and $s(t)=0$, represent the sleeping and contention stages, respectively. The stages $s(t)=\{1, 2, \cdots, R\}$ represent transmission stages corresponds to repetition rate $R$. The stage $s(t) = R+1$ represents the waiting stage of the device after all its unsuccessful transmissions. Finally, $s(t) = R+2$ represents the success stage of the device. Moreover, the number of states in each transmission stage is equal to $T-R+1$ and the number of states in the waiting stage is equal to $T-R$.

The bidimensional process $\{s(t),b(t)\}$ of the proposed mechanism is modeled using a discrete-time Markov chain for a repetition rate $R=2$ as depicted in Fig.~\ref{Mod12}. Let $s_{i,j}$ represents a state in the Markov chain where, $s(t) = i$ and $b(t)=j$. 
The transition probability from $s_{-1,0}$ to $s_{0,j}$ for each $j\in \{0, T-1\}$ is given as 
\begin{equation*}
    P(0,j|-1,0)=  \frac{1-P_{Sleep}}{T}\,, 
\end{equation*}
where, $P_{Sleep}$ defines the probability that a device has no packet to transmit and remains in the sleep state.
%
%
%doesn't get a packet to transmit.
%
%{\color{red}which accounts for the fact that a device having a packet to transmit moves from sleeping stage to contention stage with a probability $1-P_{Sleep}$ and selects one state in contention stage.}

Let $L$ represents the average number of devices that are contending at $s_{0,0}$.
Then, a device in $s_{0,0}$ enters into the first transmission stage $s_{1,j}$ $\forall$ $j\in \{0,T-2\}$  with a probability $P$ given as
\begin{align} \label{pro_ent}
P= \frac{E\,K\,T}{L}\,,
\end{align}
else, 
it remains in the contention cycle with a probability $(1-P)$ till the beginning of the next radio frame cycle. 
Here, $P$ is chosen in such a way that on an average $E\,K\,T$ devices enter into the first transmission stage.
With $K$ being the number of orthogonal preambles, the average number of devices with identical preamble in the first transmission stage is given as
\begin{equation*}
N = \frac{E\,K\,T}{K} = {E\,T\,,}
\end{equation*} 
where, $E$ and $T$ are as explained earlier in Section~\ref{proposed_algo}.
\subsection{Transition Probabilities}
In this subsection, we derive the expressions for transition probabilities $\alpha_j$, $\beta_j$, and $\gamma_j$ 
and transmission probabilities $P_{t1}$ and $P_{t2}$, as shown in Fig.~\ref{Mod12},
as a function of $T$ and $N$.
Let $r_1$ and $r_2$ be two dissimilar slots a device selects uniformly from the set $S=\{0,T-1\}$ with $R=2$. Thus,
the probabilities corresponding to the selection of $r_1$, followed by $r_2$ are derived as
\begin{align*}
    \textbf{Pr}(r_1)= 
\begin{dcases}
    \frac{1}{T},& \,\, r_1\in S\,;  \\
    0,              & \text{otherwise}\,,
\end{dcases}
\end{align*}
\begin{align*}
    \textbf{Pr}(r_2|r_1)= 
\begin{dcases}
    \frac{1}{T-1},&  r_2\in S\backslash \{r_1\}\,;  \\
    0,              & \text{otherwise}\,,
\end{dcases}
\end{align*}
respectively.

Let $r_{\min}=\min\{r_1,r_2\}$ represents the remaining time slots before the first transmission. Thus, in Fig.~\ref{Mod12}, $\alpha_j$ represents the probability that a device has to wait for $r_{\min}=j$ time slots before the first transmission. Hence, $\alpha_j$ for each $j\in \{0,T-2\}$ can be derived as   
\begin{align*}
\alpha_j &= \textbf{Pr}[r_{\min}=j] \nonumber \\ \nonumber
&=\textbf{Pr}[r_1 = j,r_2 > j]+\textbf{Pr}[r_1 > j, r_2 = j] \nonumber \\
&= \frac{1}{T}\frac{T-j-1}{T-1} + \frac{1}{T-1}\frac{T-j-1}{T} \nonumber \\
&=\frac{2(T-j-1)}{T(T-1)}\,. 
\end{align*}
A device in $s_{1,j}$ for each $j\in \{0,T-2\}$ waits for $j$ time slots and transmits with a success probability $P_{t1}$ given as~\cite{Casini}
\begin{align} \label{P_t1}
P_{t1} =\left (1-\frac{2}{T}\right)^{N-1}\,.
\end{align}
In case the device is unsuccessful, it moves to second transmission stage 
$s_{2,j}$ $\forall$ $j \in \{0,T-2\}$ with a probability $(1-P_{t1})$. 

Let $r_{\max}=\max\{r_1,r_2\}$, then, 
$(r_{\max}-r_{\min}-1)$ will be the remaining time slots a device has to wait after the first transmission and before the second transmission. Let $\beta_j$ represents the probability that a device has to wait for $(r_{\max}-r_{\min}-1)=j$ time slots after the first transmission and before the second transmission. Then, $\beta_j$ for each $j\in \{0,T-2\}$ can be derived as 
\begin{align} 
\beta_j &= \textbf{Pr}[r_{\max}-r_{\min}-1=j] \nonumber \\ \nonumber
&=\left\{\textbf{Pr}\left[r_2 = r_1+(j +1)|r_1\right]\right\}\textbf{Pr}[r_1] \\
&\hspace{0.5cm}+ \left\{ \textbf{Pr}\left[r_2 = r_1 - (j+1) |  r_1\right]\right\}\textbf{Pr}[r_1] \nonumber \\
&= \frac{2(T-2(j+1)) + 2(j+1)}{T(T-1)} \nonumber \\
&= \frac{2(T-j-1)}{T(T-1)}\,. \label{gamma_j}
\end{align}
Let $P_{SIC}(i,j)$ for $i\in \{2,3\}$ and $j\in \{0,T-2\}$ be the probability of SIC  at $s_{i,j}$. Then, a device in $s_{2,j}$ for each $j \in \{1,T-2\}$ gets success with $P_{SIC}(2,j)$,  else, it moves to $s_{2,j-1}$ with probability $\left(1-P_{SIC}(2,j)\right)$. The detailed derivation of $P_{SIC}(2,j)$ is given in the next subsection.
A device that is unsuccessful with SIC in $s_{2,1}$ moves to $s_{2,0}$ and transmits with a success probability $P_{t2}$ derived as
\begin{align} \label{P_t2}
P_{t2} = \left (1-\frac{2}{T}\right)^{\psi N-1}\,.
\end{align}
In case the device is unsuccessful, it moves to waiting stage $s_{3,j}$ $\forall$ $j\in \{0,T-3\}$ with a probability $(1-P_{t2})$.
Here,
\begin{align*}
\psi = \left(1-P_{t1}\right)\left(\beta_0 + \sum_{i=1}^{T-2} \prod_{j=1}^{i} \left\{1-P_{SIC}(2,j)\right\}\beta_{i}\right)
\end{align*}
is the probability that a device is unsuccessful in first transmission as well as with SIC and reaches to $s_{2,0}$.

Let $(T-r_{\max}-2)$ be the remaining time slots after an unsuccessful transmission at $s_{2,0}$. Thus, $\gamma_j$ represents the probability that a device has to wait for $(T-r_{\max}-2)=j$ time slots until the current radio frame cycle finishes. Hence, $\gamma_{j}$ for each $j\in \{0,T-3\}$ can be derived as
\begin{align}\label{delta_j}
\gamma_{j} &= \textbf{Pr}[T-r_{\max}-2=j] \nonumber\\ \nonumber
&= \textbf{Pr}[r_{\max}=T-j-2] \\
&=\textbf{Pr}[r_1 = T-j-2,r_2 < T-j-2] \nonumber \\&\hspace{0.4cm}+\textbf{Pr}[r_1 < T-j-2, r_2 = T-j-2] \nonumber \\
&= \frac{1}{T}\frac{T-j-2}{T-1} + \frac{1}{T-1}\frac{T-j-2}{T} \nonumber \\
&= \frac{2(T-j-2)}{T(T-1)}\,. 
\end{align}
A device in $s_{3,j}$ for each $j\in \{1,T-3\}$ gets success with probability $P_{SIC}(3,j)$, else, moves to $s_{3,j-1}$ with a probability $\left(1-P_{SIC}(3,j)\right)$. Finally, the device in $s_{3,0}$ gets success with a probability $P_{SIC}(3,0)$, else, moves to sleeping state $s_{-1,0}$ with a probability $\left(1-P_{SIC}(3,0)\right)$. 
The expression for $P_{SIC}(i,j)$ for $i=2$ and $3$ is derived in the following subsection.
\subsection{Probability of Successive Interference Cancellation}
In this subsection, we derive the probability of
SIC $\left(P_{SIC}(i,j)\right)$ for a given state $s_{i,j}$ for $i=\{2,3\}$ and $j\in \{0, 1, \cdots, T-2\}$ as depicted in Fig.~\ref{Mod12}. 
Here, $P_{SIC}(2,j)$ is the probability of SIC at $s_{2,j}$ post an unsuccessful transmission in the first attempt at $s_{1,0}$ and prior to the second transmission at $s_{2,0}$. Similarly, $P_{SIC}(3,j)$ represents the probability of SIC at $s_{3,j}$ following 
both unsuccessful transmissions at $s_{1,0}$ and $s_{2,0}$.

Let $\Gamma_2$ defines the probability that a device with $r_{\min}$ and $r_{\max}$ transit to $s_{2,j}$ where $j=(r_{\max}-r_{\min}-1)$ after an unsuccessful transmission at $s_{1,0}$ according to (\ref{gamma_j}).
A device at $s_{1,0}$ is considered to be unsuccessful if it shares an identical preamble with atleast one more device during its first transmission in time slot $r_{\min}$. Thus, $\Gamma_2$ can be expressed as
\begin{align}
\Gamma_2&= \sum_{i=1}^{N-1} \binom{N-1}{i}\left(\frac{2}{T}\right)^i \left(1-\frac{2}{T}\right)^{N-1-i}\,, \nonumber \\
&= 1-\left(1-\frac{2}{T}\right)^{N-1}\,. \label{c_1} 
\end{align}

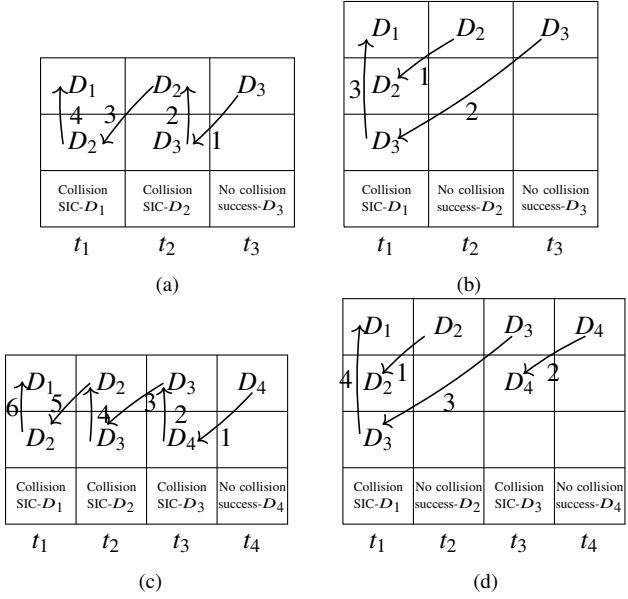
\begin{figure}[!tbp]
   \centering
   \subfloat[][]{
  \begin{tikzpicture}[x=.75cm,y=.5cm]
  \draw (0,0) -- (4.5,0) -- (4.5,4.5) -- (0,4.5) --(0,0);
\draw (0,1.5) -- (4.5,1.5);
\draw (0,3) -- (4.5,3);
\draw (1.5,0) -- (1.5,4.5);
\draw (3,0) -- (3,4.5); 
\node at (0.75,3.75) [](d1) {$D_1$};
\node at (2.25,3.75) [] (d2){$D_2$};
\node at (3.75,3.75) [](d3) {$D_3$};
\node at (2.25,2.25) [] (d4){$D_3$};
\node at (0.75,2.25) [] (d5){$D_2$};

\node at (0.75,-0.5) [](d1) {$t_1$};
\node at (2.25,-0.5) [] (d2){$t_2$};
\node at (3.75,-0.5) [](d3) {$t_3$};
%
%{\scriptsize{Select a number m from a uniformly} \\ \scriptsize{distributed set $\{0, 1, \cdots, T-1\}$}};

\node at (3.75,1) [align=center]  {\tiny{No collision}};

 \node at (3.75,0.5) [align=center] {\tiny{success-$D_3$}};
 
 \node at (2.25,1) [align=center]  {\tiny{Collision}};

 \node at (2.25,0.5) [align=center] {\tiny{SIC-$D_2$}};
 
 \node at (0.75,1) [align=center]  {\tiny{Collision}};

 \node at (0.75,0.5) [align=center] {\tiny{SIC-$D_1$}}; 
 
%\node at (2.25,0.75) [] {\small{SIC-$D_2$}};
%\node at (0.75,0.75) [] {\small{SIC-$D_1$}};

%\node at (2.25,-0.75) [] {\small{(a)}} ;
%\node at (7.25,-0.75) [] {\small{(b)}} ;

\path[->,semithick] (3.5,3.5)   edge[bend left =5,anchor=center] node[below] {1}         (2.7,2.2)

(2.6,2.2)   edge[bend right =5,anchor=center] node[left] {2}         (2.6,3.75)

(2.0,3.75)   edge[bend right =5,anchor=center] node[left] {3}         (1.1,2.2)

(0.4,2.2)   edge[bend left =5,anchor=center] node[right] {4}         (0.35,3.75);

\end{tikzpicture}  
  \label{i=3a}}\quad
   \subfloat[][]{
\begin{tikzpicture}  [x=.75cm,y=.5cm]
  \draw (5,0) -- (9.5,0) -- (9.5,6) -- (5,6) --(5,0);
\draw (5,1.5) -- (9.5,1.5);
\draw (5,3) -- (9.5,3);
\draw (5,4.5) -- (9.5,4.5);
\draw (6.5,0) -- (6.5,6);
\draw (8,0) -- (8,6);

\node at (5.75,5.25) [](d1) {$D_1$};
\node at (7.25,5.25) [] (d2){$D_2$};
\node at (8.75,5.25) [](d3) {$D_3$};
%\node at (2.25,2.25) [] (d4){$D_3$};
\node at (5.75,3.75) [] (d5){$D_2$};
\node at (5.75,2.25) [] (d5){$D_3$};

\node at (5.75,-0.5) [](d1) {$t_1$};
\node at (7.25,-0.5) [] (d2){$t_2$};
\node at (8.75,-0.5) [](d3) {$t_3$};

\node at (8.75,1) [align=center]  {\tiny{No collision}};

 \node at (8.75,0.5) [align=center] {\tiny{success-$D_3$}};
 
 \node at (7.25,1) [align=center]  {\tiny{No collision}};

 \node at (7.25,0.5) [align=center] {\tiny{success-$D_2$}};
 
 \node at (5.75,1) [align=center]  {\tiny{Collision}};

 \node at (5.75,0.5) [align=center] {\tiny{SIC-$D_1$}};

%\node at (8.75,0.75) [] {\small{Txion}};
%\node at (7.25,0.75) [] {\small{Txion}};
%\node at (5.75,0.75) [] {\small{SIC-$D_1$}};
%\node at (3.75,0.75) [] {\small{Txion}};
%\node at (2.25,0.75) [] {\small{SIC-$D_2$}};
%\node at (0.75,0.75) [] {\small{SIC-$D_1$}};
%\draw[->] (2.5,2.25) -- (1.77,1.5)node[bend left  =3cm,anchor=center, yshift=0.15cm]{};
\path[->,semithick] (8.5,5)   edge[bend left =5,anchor=center] node[below] {2}         (5.95,2.35)

(5.4,2.35)   edge[bend left =5,anchor=center] node[left,xshift=0.1cm] {3}         (5.4,5)

(6.95,5)   edge[bend right =5,anchor=center] node[below] {1}         (5.950,3.95);

\end{tikzpicture}
\label{i=3b}}\\
   \subfloat[][]{\begin{tikzpicture}[x=.75cm,y=.5cm]
\draw (0,0) -- (5,0) -- (5,4.5) -- (0,4.5) --(0,0);
\draw (0,1.5) -- (5,1.5);
\draw (0,3) -- (5,3);
%\draw (0,4.5) -- (5,4.5);
\draw (1.25,0) -- (1.25,4.5);
\draw (2.5,0) -- (2.5,4.5);
\draw (3.75,0) -- (3.75,4.5);
%\draw (3,0) -- (3,4.5); 
\node at (0.625,3.75) [](d1) {$D_1$};
\node at (1.875,3.75) [](d2) {$D_2$};
\node at (3.125,3.75) [] (d3){$D_3$};
\node at (4.375,3.75) [](d4) {$D_4$};

\node at (0.625,2.25) [] (d5){$D_2$};
\node at (1.875,2.25) [] (d6){$D_3$};
\node at (3.125,2.25) [] (d7){$D_4$};

\node at (0.625,-0.5) [](d1) {$t_1$};
\node at (1.875,-0.5) [](d2) {$t_2$};
\node at (3.125,-0.5) [] (d3){$t_3$};
\node at (4.375,-0.5) [](d4) {$t_4$};

\node at (4.375,1) [align=center]  {\tiny{No collision}};

 \node at (4.375,0.5) [align=center] {\tiny{success-$D_4$}};
 
 \node at (3.125,1) [align=center]  {\tiny{Collision}};

 \node at (3.125,0.5) [align=center] {\tiny{SIC-$D_3$}};
 
 \node at (1.875,1) [align=center]  {\tiny{Collision}};

 \node at (1.875,0.5) [align=center] {\tiny{SIC-$D_2$}}; 
 
  \node at (0.625,1) [align=center]  {\tiny{Collision}};

 \node at (0.625,0.5) [align=center] {\tiny{SIC-$D_1$}};

%\node at (4.375,0.75) [] {\small{Txion}};
%\node at (3.125,0.75) [] {\small{SIC-$D_3$}};
%\node at (0.625,0.75) [] {\small{SIC-$D_1$}};
%\node at (1.875,0.75) [] {\small{SIC-$D_2$}};

\path[->,semithick] (4.375,3.5)   edge[bend left =5,anchor=center] node[below] {1}         (3.4,2.2)

(2.8,2.2)   edge[bend right =5,anchor=center] node[right] {2}         (2.8,3.65)

(2.8,3.75)   edge[bend right =5,anchor=center] node[right] {3}         (1.8,2.65)

(1.5,2.2)   edge[bend right =5,anchor=center] node[right=-0.05cm] {4}         (1.5,3.65)

(1.5,3.75)   edge[bend right =5,anchor=center] node[left = -0.05cm] {5}         (0.8,2.65)

(0.3,2.45)   edge[bend left =5,anchor=center] node[left=-0.1cm] {6}         (0.3,3.8);
\end{tikzpicture}
\label{i=4a}}\quad
   \subfloat[][]{\begin{tikzpicture}[x=.75cm,y=.5cm]
  
\draw (5.5,0) -- (10.5,0) -- (10.5,6) -- (5.5,6) --(5.5,0);
\draw (5.5,1.5) -- (10.5,1.5);
\draw (5.5,3) -- (10.5,3);
\draw (5.5,4.5) -- (10.5,4.5);

\draw (6.75,0) -- (6.75,6);
\draw (8,0) -- (8,6);
\draw (9.25,0) -- (9.25,6);
%\draw (8,0) -- (8,6);

\node at (6.125,5.25) []() {$D_1$};
\node at (7.375,5.25) [] (d2){$D_2$};
\node at (8.625,5.25) [](d3) {$D_3$};
\node at (9.875,5.25) [](d3) {$D_4$};

\node at (6.125,-0.5) [](d1) {$t_1$};
\node at (7.375,-0.5) [](d2) {$t_2$};
\node at (8.625,-0.5) [] (d3){$t_3$};
\node at (9.875,-0.5) [](d4) {$t_4$};

\node at (6.125,3.75) [](d1) {$D_2$};
%\node at (7.375,3.75) [] (d2){$D_4$};
\node at (8.625,3.75) [](d3) {$D_4$};
%\node at (9.875,3.75) [](d3) {$D_4$};

\node at (6.125,2.25) [](d1) {$D_3$};
%\node at (7.375,2.5) [] (d2){$D_2$};
%\node at (8.625,2.5) [](d3) {$D_3$};
%\node at (9.875,2.5) [](d3) {$D_4$};

%\node at (2.25,2.25) [] (d4){$D_3$};
%\node at (5.75,3.75) [] (d5){$D_2$};
%\node at (5.75,2.25) [] (d5){$D_3$};

\node at (9.875,1) [align=center]  {\tiny{No collision}};

 \node at (9.875,0.5) [align=center] {\tiny{success-$D_4$}};
 
 \node at (8.625,1) [align=center]  {\tiny{Collision}};

 \node at (8.625,0.5) [align=center] {\tiny{SIC-$D_3$}};
 
 \node at (7.375,1) [align=center]  {\tiny{No collision}};

 \node at (7.375,0.5) [align=center] {\tiny{success-$D_2$}}; 
 
  \node at (6.125,1) [align=center]  {\tiny{Collision}};

 \node at (6.125,0.5) [align=center] {\tiny{SIC-$D_1$}};

%\node at (9.875,0.75) [] {\small{Txion}};
%\node at (8.625,0.75) [] {\small{SIC-$D_3$}};
%\node at (7.375,0.75) [] {\small{Txion}};
%\node at (6.125,0.75) [] {\small{SIC-$D_1$}};
%\node at (3.75,0.75) [] {\small{Txion}};
%\node at (2.25,0.75) [] {\small{SIC-$D_2$}};
%\node at (0.75,0.75) [] {\small{SIC-$D_1$}};
%\draw[->] (2.5,2.25) -- (1.77,1.5)node[bend left  =3cm,anchor=center, yshift=0.15cm]{};
\path[->,semithick] (8.5,5)   edge[bend left =5,anchor=center] node[below] {3}         (6.2,2.65)

(6.95,5)   edge[bend right =5,anchor=center] node[below] {1}         (6.2,4)

(9.8,5)   edge[bend right =5,anchor=center] node[below] {2}         (8.7,4)

(5.8,2.4)   edge[bend left =5,anchor=center] node[left=-0.07cm] {4}         (5.8,5.3);

%\draw (0,-2)--(0,-9.5) -- (5,-9.5) -- (5,-2) -- (0,-2);
%
%\draw (1.25,-9.5) -- (1.25,-2);
%\draw (2.5,-9.5) -- (2.5,-2);
%\draw (3.75,-9.5) -- (3.75,-2);
%
%\draw (0,-3.5) -- (5,-3.5);
%\draw (0,-5) -- (5,-5);	
%\draw (0,-6.5) -- (5,-6.5);
%\draw (0,-8) -- (5,-8);
%
%\node at (0.625,-2.75) [](d1) {$D_1$};
%\node at (1.875,-2.75) [](d2) {$D_2$};
%\node at (3.125,-2.75) [] (d3){$D_3$};
%\node at (4.375,-2.75) [](d4) {$D_4$};
%
%
%\node at (0.625,-4.25) [](d1) {$D_2$};
%\node at (0.625,-5.75) [](d1) {$D_3$};
%\node at (0.625,-7.25) [](d1) {$D_4$};
%
%\node at (4.375,-8.75) [] {\small{Txion}};
%\node at (3.125,-8.75) [] {\small{Txion}};
%\node at (0.625,-8.75) [] {\small{SIC-$D_1$}};
%\node at (1.875,-8.75) [] {\small{Txion}};
%
%\path[->,semithick] (4.375,-2.95)   edge[bend left =10,anchor=center] node[below] {3}         (0.9,-7.25)
%
%
%(3.125,-2.95)   edge[bend left =10,anchor=center] node[below] {2}         (0.9,-5.75)
%
%(1.875,-2.95)   edge[bend left =10,anchor=center] node[below] {1}         (0.9,-4.35)
%
%(0.225,-7.25)   edge[left =5,anchor=center] node[left=-0.07cm] {4}         (0.225,-2.75);
%\node at (2.5,-10.25) [](d4) {\small{(c)}};
\end{tikzpicture}  
  
  \label{i=4b}}
   \caption{(a) and (b) show a set of two possible arrangements of three devices in three time slots such that $D_1$ (the device of interest) gets success with SIC and (c) and (d) show a set of two possible arrangements of $4$ devices in $4$ time slots such that $D_1$ gets success with SIC.}
   \label{arrang}
\end{figure}

Let $i$ devices, with identical preamble, are distributed over $i$ time slots such that there is a successful transmission at last time slot. Let one of the devices has chosen $r_{\min}$ from first $i-1$ time slots. Thus, $\delta_i$ defines the probability that the same device gets success with SIC at $s_{2,j}$ where $j=(r_{\max}-r_{\min}-i)$ due to successful transmissions and back-and-forth SIC from other devices. 
Figs.~\ref{i=3a} and \ref{i=3b} show two possible arrangements correspond to $i=3$.
Here, $D_1$, $D_2$, and $D_3$ represent three devices and $t_1$, $t_2$, and $t_3$ represent three time slots. Hence, $\delta_3$ can be obtained as 
\begin{align} \label{del_3}
\delta_3 = \binom{N-1}{2}2^2\left(\frac{1}{\binom{T}{2}}\right)^{2}  \left(\frac{\binom{T-3}{2}}{\binom{T}{2}}\right)^{N-3} \,.
\end{align} 
Similarly, from Figs.~\ref{i=4a} and \ref{i=4b}, the expression for $\delta_4$ can be obtained as  
\begin{align}\label{del_4}
\delta_4 = \binom{N-1}{3}3^2 2^2\left(\frac{1}{\binom{T}{2}}\right)^{3}  \left(\frac{\binom{T-4}{2}}{\binom{T}{2}}\right)^{N-4}\,.
\end{align}
From (\ref{del_3}) and (\ref{del_4}), the generalized expression for $\delta_i$ for each $i\in \{2,T-2\}$ is obtained as
\begin{align}
\delta_i &=  \binom{N-1}{i-1}\left(\frac{1}{\binom{T}{2}}\right)^{i-1} \prod_{j=1}^{i-1} (i-j)^2\left(\frac{\binom{T-i}{2}}{\binom{T}{2}}\right)^{N-i}\,, \nonumber \\
&=\frac{i}{NP_\delta^i}\left(\frac{1}{\binom{T}{2}}\right)^{i-1}((i-1)!)^2  \binom{N}{i} P_\delta^i \,(1-P_\delta)^{N-i}, \label{alpha_i}
\end{align}
%and we can re-write the above expression as
%\begin{align}
%\delta_i=\frac{i}{NP_\delta^i}\left(\frac{1}{\binom{T}{2}}\right)^{i-1}((i-1)!)^2  \binom{N}{i} P_\delta^i \,(1-P_\delta)^{N-i}, \label{alpha_i}
%\end{align}
where, $P_{\delta_i}\! = \! 1\!-\!\frac{\binom{T-i}{2}}{\binom{T}{2}}$.
By extending the arrangements in Fig.~\ref{arrang}, $C^{(2)}_{j,k}$ defines the total number of combinations obtained from all possible arrangements of $k$ devices in $j$ time slots for a successful SIC at $s_{2,T-j}$ and is obtained as 
\begin{align}
C^{(2)}_{j,k}
& = \sum_{l=k}^{T-j} \left[\sum_{i=1}^{k-1}\left\{ \binom{l-1}{1}\binom{l-2}{k-2} (k\!-\!1\!-\!i) \right\}\!+\! \binom{l\!-\!1}{1} \binom{l-2}{k-2}\right]\,,\nonumber \\
&= \sum_{l=k}^{T-j} \left[ \binom{l-1}{1}\binom{l-2}{k-2}\left\{\frac{(k-2)(k-1)}{2} +1\right\}\right]\,. \label{C_JK}
\end{align}
From (\ref{c_1}), (\ref{alpha_i}), and (\ref{C_JK}), $P_{SIC}(2,j)$ can be derived as 
 \begin{align} \label{P_SIC}
P_{SIC}(2,j) 
 = \begin{cases}
 \frac{\sum\limits_{k=2}^{T-j}C^{(2)}_{j,k}\,\delta_k}{\left(C^{(2)}_{j,2}-\sum\limits_{u=j+1}^{T-2}C^{(2)}_{u,2}\,P_{SIC}(2,u)\right)\Gamma_2}, & \!\! j\in \{1,T-2\}\,; \\
 0, &\!\! \text{otherwise}\,,
 \end{cases}
\end{align} 
where, the denominator denotes the total number of unsuccessful devices that reach $s_{2,j}$
after either an immediate unsuccessful transmission at $s_{1,0}$ (and transit with $\beta_j$) or an unsuccessful SIC in the higher state $s_{2,j+1}$ $\left(\text{and transit with } 1-P_{SIC}(2,j+1)\right)$ as shown in Fig.~\ref{Mod12}. Further, the numerator defines the total number of 
possible combinations 
obtain from the arrangement of $2$ to $T-j$ devices in $T-j$ slots such that a device gets success with SIC at $s_{2,j}$. 
Similarly, the expression for $P_{SIC}(3,j)$ is derived as
\begin{align} \label{P_wt}
P_{SIC}(3,T-j) =\begin{cases}
 \frac{\sum\limits_{k=2}^{j-1}C^{(3)}_{j,k}\,\delta_k}{\left(\frac{C^{(3)}_{j,2}}{2}-\sum\limits_{u=1}^{j-1} \frac	{C^{(3)}_{u,2}}{2}\,P_{SIC}(3,u)\right)\Gamma_3}, &  j \in \{\ 3,T\}\,; \\ 
 0, & \text{otherwise}\,,
 \end{cases}
\end{align}
where, the denominator denotes the total number of unsuccessful devices that reach $s_{3,T-j}$
after either an immediate unsuccessful transmission at $s_{2,0}$ (and transit with $\gamma_{T-j}$) or an unsuccessful SIC in the higher state $s_{3,T-j+1}$ (and transit with $1-P_{SIC}(3,T-j+1)$) as shown in Fig.~\ref{Mod12}. Further, the numerator defines the total number of 
possible combinations 
obtain from the arrangement of $2$ to $j-1$ devices in $j$ time slots such that a device gets success with SIC at $s_{3,j}$.
Here, $C_{j,k}^{(3)}$ defines the total number of combinations obtained from all possible arrangements of $k$ devices in $j$ time slots for a successful SIC at $s_{3,T-j}$ and is obtained as
\begin{align}
C^{(3)}_{j,k} =  \binom{2}{1} \binom{l-3}{k-2}  \left\{\frac{(k-2)(k-1)}{2} +1\right\} \,. \label{d}
\end{align}
Moreover, $\Gamma_3$ defines the probability that a device with $r_{\max}$ transits to $s_{3,j}$, where, $j=(T-r_{\max}-2)$ after an unsuccessful transmission during both transmissions at  $s_{1,0}$ and $s_{2,0}$ according to (\ref{delta_j}). 
A device is considered as unsuccessful at $s_{1,0}$ and $s_{2,0}$ if it shares an identical preamble with atleast one more device in time slots $r_{\min}$ and $r_{\max}$, respectively.
Thus, $\Gamma_3$ can be expressed as
\begin{align*}
\Gamma_3 &= \sum_{i=2}^{N} \binom{N}{i} \left(\frac{1}{\binom{T}{2}}\right) p_{\tau}^{(i-1)} (1-p_{\tau})^{(N-i)}\,,  \nonumber \\
&= \left(\frac{1}{\binom{T}{2}p_{\tau}}\!\right)\left[1-(1+(N-1)p_{\tau})(1-p_{\tau})^{N-1})\right]\,,  
\end{align*} 
where, $p_{\tau} =	1-\frac{\binom{T-2}{2}}{\binom{T}{2}}$. Next, we present the steady-state analysis for the Markov chain presented in Fig.~\ref{Mod12}.
\section{steady-state Analysis} \label{STD}
Let $b_{i,j}=\lim\limits_{t\to\infty} \textbf{Pr}\{s(t) = i, b(t) = j\}$ $\forall$ $i\in \{-1,4\}$ and $j \in \{0,T-1\}$ be the stationary distribution of the Markov chain. Then, from Fig.~\ref{Mod12}, the steady-state probabilities $b_{0,0}$ and $b_{1,0}$ are obtained as 
\begin{align*}
b_{0,0} = \frac{1-P_{Sleep}}{P}b_{0,-1},\text{ and}
\end{align*}
\begin{align} \label{b1k}
b_{1,0} = P\,b_{0,0}\,,
\end{align}
respectively.

A device that is unsuccessful in first transmission at $s_{1,0}$ as well as with SIC at $s_{2,j+1}$ reaches to $s_{2,j}$. Thus, from Fig.~\ref{Mod12}, $b_{2,j}$ for each $j\in \{0,T-2\}$ is obtained as
\begin{align} \label{b2k}
b_{2,j}  &= \left(1-P_{t1}\right)\left[\beta_j + \sum_{k=j+1}^{T-2} \prod_{l=j+1}^{k} \left(1-P_{SIC}(2,l)\right)\beta_{l} \right]b_{1,0} \nonumber \\
&=P\left(1-P_{t1}\right)\left[\beta_j + \sum_{k=j+1}^{T-2} \prod_{l=j+1}^{k} \left(1-P_{SIC}(2,l)\right)\beta_{l} \right]b_{0,0}\,.
\end{align}
Similarly, a device that is unsuccessful in second transmission at $s_{2,0}$ as well as with SIC at $s_{3,j+1}$ reaches to $s_{3,j}$. Thus, from Fig.~\ref{Mod12}, $b_{3,j}$ for each $j\in \{0,T-3\}$ is obtained as
\begin{align} \label{b3k}
b_{3,j} &= \left(1-P_{t2}\right)\left[\!\gamma_j + \sum_{k=j+1}^{T-3} \prod_{l=j+1}^{k} \left(1-P_{SIC}(3,l)\right)\gamma_{l} \right]b_{2,0} \nonumber \\ 
&=P\left(1-P_{t1}\right)\left[\beta_0 + \sum_{k=1}^{T-2} \prod_{l=1}^{k} \left(1-P_{SIC}(2,l)\right)\beta_{l} \right] \nonumber  \\
 &\hspace{0.3cm}\times \left(1-P_{t2}\right)\left[\gamma_j + \sum_{k=j+1}^{T-3} \prod_{l=j+1}^{k} \left(1-P_{SIC}(3,l)\right)\gamma_{l} \right]\!b_{0,0} .
\end{align}
A device that is successful either 
during transmission at $s_{1,0} $ or $s_{2,0}$ or with SIC at $s_{2,j}$ or $s_{3,j}$ 
using (\ref{P_t1}), (\ref{P_t2}), (\ref{P_SIC}), (\ref{P_wt}), (\ref{b1k}), (\ref{b2k}), and (\ref{b3k})
reaches to $s_{4,0}$. Thus, $b_{4,0}$ is obtained as
\begin{align} \label{std_b30}
b_{4,0} =& P_{t1}b_{1,0} + P_{t2}b_{2,0}+ \sum_{l=1}^{T-2}P_{SIC}(2,l)b_{2,l} \nonumber \\ & + \sum_{l=0}^{T-3}P_{SIC}(3,l)b_{3,l}\,. 
\end{align}

\section{numerical results} \label{RES}
Numerical results along with the simulation results are presented in this section to validate the analytical model. 
We also comment on how the value of various parameters affects the throughput of the proposed mechanism. Further, we compare the success rate of the proposed mechanism with the existing ones.
\subsection{Validation of the Proposed Analytical Model}
In this subsection, we validate the analytical model for the proposed SIC based RACH mechanism using numerical and simulation results. Wherein, all the results are obtained from the MATLAB using Monte-Carlo simulation. Since the main focus of the paper is on the RACH mechanism, we do not consider node mobility.
We consider an average of $0.1$ million devices at state $(0,0)$ in the beginning of each radio frame cycle. 
Thus, the value of $L$ and $\cal{C}$ are $10^5$ and $10^{-5}$, respectively.
We consider the radio frame cycle length, $T$, in the integer range of $100-1000$ slots and the number of preambles, $K$, in the range  of $10-70$. 
Further, we select the fractional $E$ between $0$ and $1$ and set the repetition rate at $R=2$. 

\begin{figure}[!tbp]
  \centering
  \subfloat[]{\includegraphics[width=3.45 in, height=2.5 in]{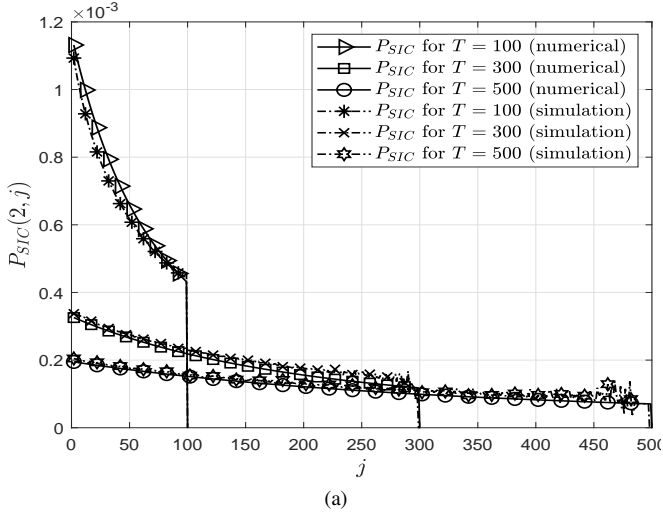}\label{psicfig}}
  \hfill
  \subfloat[]{\includegraphics[width=3.45 in, height=2.5 in]{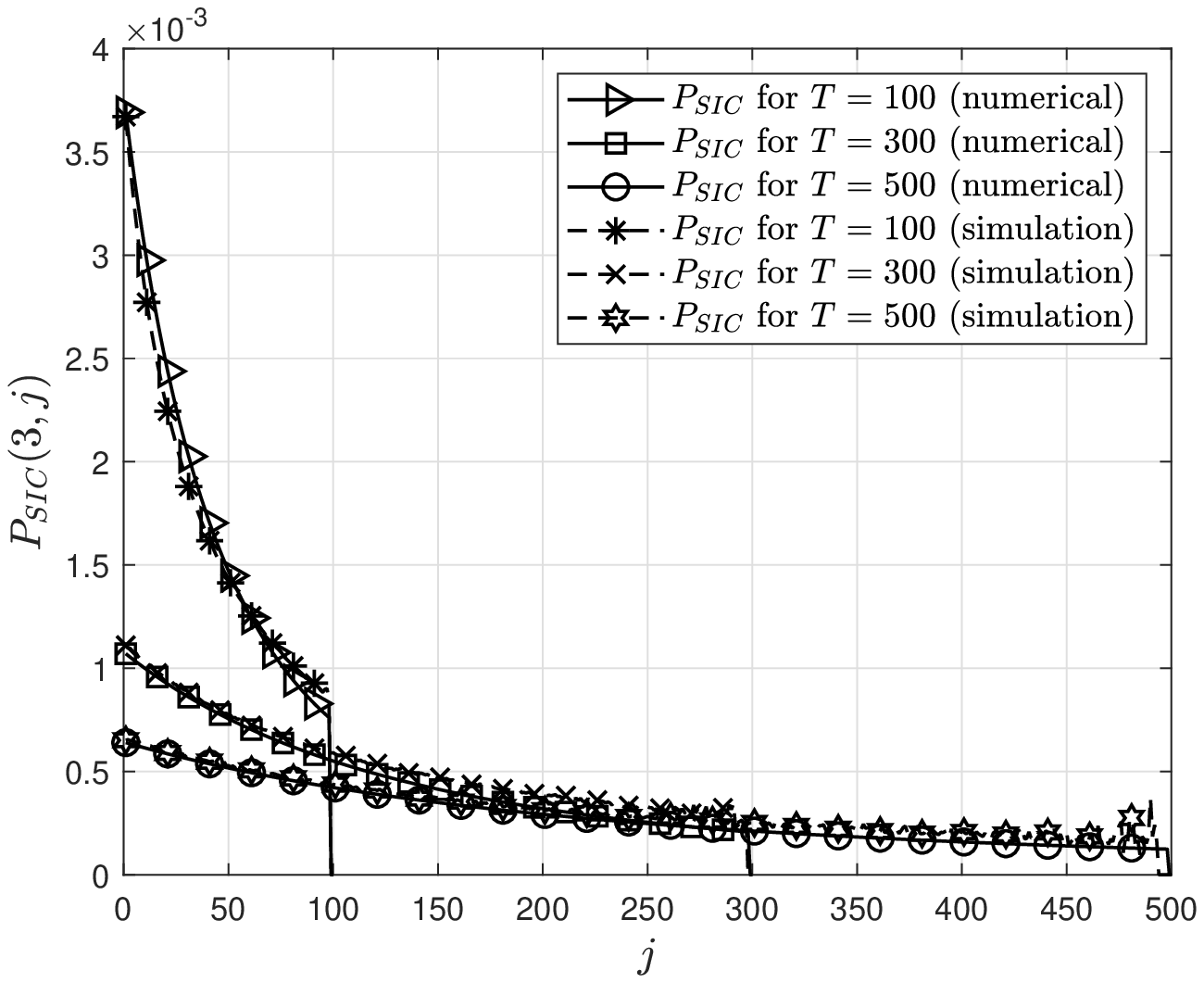}\label{pwtfig}}
  \caption{$P_{SIC}$ for different values of T at (a) state $(2,j)$ for $j\in \{1, 2, \cdots, T-2\}$ and (b) state $(3,j)$ for $j\in \{0, 1, \cdots, T-3\}$.}
\end{figure}

\begin{figure}[!tbp]
  \centering
  \subfloat[]{\includegraphics[width=3.45 in, height=2.5 in]{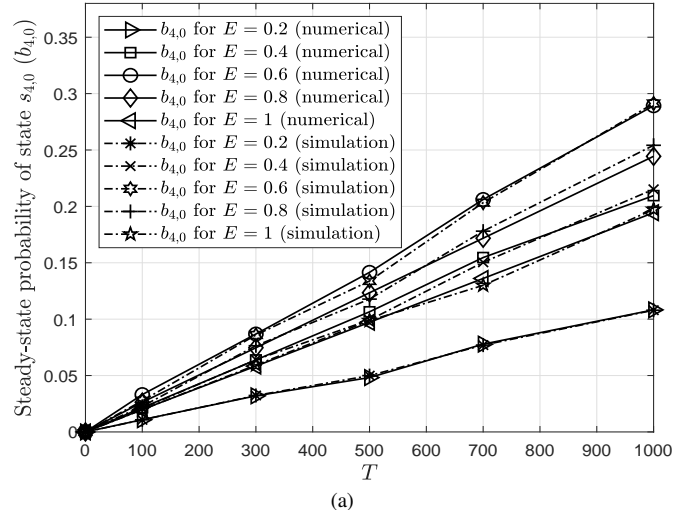}\label{b30_T}}
  \hfill
  \subfloat[]{\includegraphics[width=3.45 in, height=2.5 in]{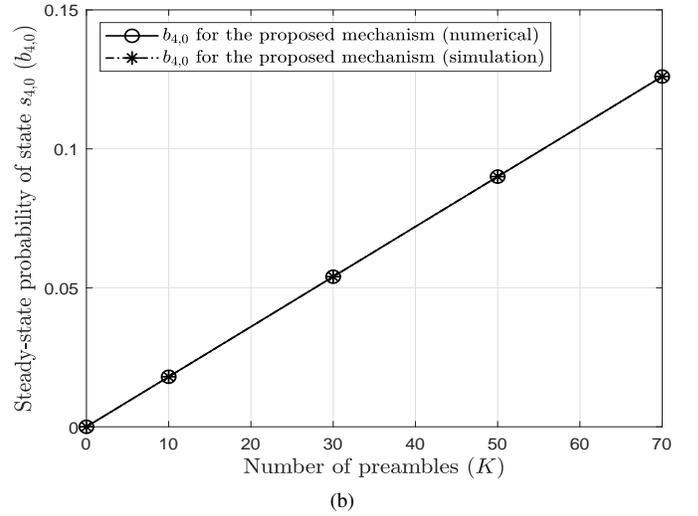}\label{b30_K}}
  \caption{Variation of steady-state probability of state $s_{4,0}$ with (a) $T$ for $K = 54$ and $E=1$ and (b) $K$ for $T=500$ and $E=1$.}
\end{figure}
For large values of $T$ ($T>100$), the binomial expression in (\ref{alpha_i}) and the binomial coefficients in  
(\ref{C_JK}) and (\ref{d}) result 
in higher values that are difficult to compute using MATLAB due to limited memory allocation.
Thus, we approximate the binomial distribution in (\ref{alpha_i}) with a normal distribution using the DeMoivre-Laplace theorem \cite{Papoulis} as
\begin{align*}
\delta_i \approx &\frac{1}{p_{\delta_i}^i\sqrt{2\pi N p_{\delta_i} q_{\delta_i}}} \exp\left((-(i-Np_\delta)^2 /{2Np_{\delta_i} q_{\delta_i}}\right)\nonumber \\ &\times \frac{i}{N} \frac{1}{\binom{T}{2}^{i-1}} ((i-1)!)^2\,.
\end{align*}
Similarly, we approximate the binomial coefficients in (\ref{C_JK}) and (\ref{d}) using Sterling's approximation \cite{sterling}. Hence, we re-express (\ref{C_JK}) and (\ref{d}) as
\begin{figure}[!tbp]
  \centering
  \subfloat[]{\includegraphics[width=3.45 in, height=2.5 in]{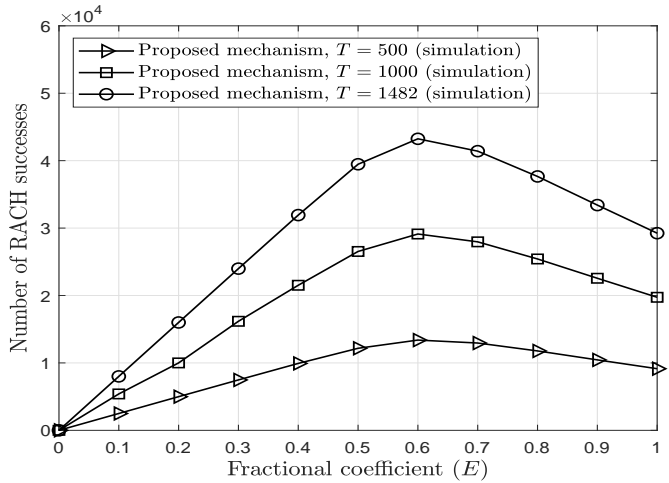}\label{maxe}}
  \hfill
  \subfloat[]{\includegraphics[width=3.45 in, height=2.5 in]{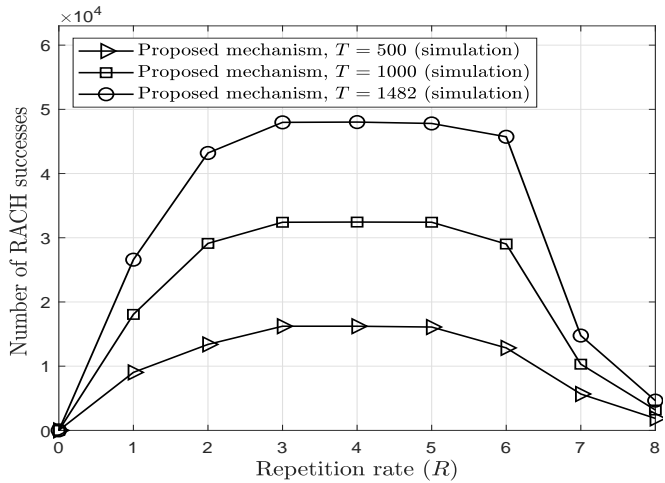}\label{compacc1}}
  \caption{Number of RACH successes made in a radio frame cycle for the proposed mechanism with respect to (a) $E$ for various values of $T$ with $k=54$ and $R=2$ and (b) $R$ for various values of $T$ with $K=54$ and $E=0.6$.}
\end{figure}
\begin{align*}
C_{j,k} &\approx  \sum_{l=k}^{T-j}\! \left[\! \!\frac{(l\!-\!1)(l\!-\!2)^{(k-2)}}{\sqrt{2\pi (k\!-\!2)}\left(\frac{k-2}{e}\right)^{ (k-2)}}\! \left\{\!\frac{(k\!-\!2)(k\!-\!1)}{2} \!+\!1\!\right\}\!\right]\,, \\
D_{j,k} &\approx  \frac{2(l\!-\!3)^{(k-2)}}{\sqrt{2\pi (k\!-\!2)}\left(\frac{k-2}{e}\right)^{ (k-2)}}\! \left\{\!\frac{(k\!-\!2)(k\!-\!1)}{2} \!+\!1\!\right\}\,.
\end{align*}

Figs.~\ref{psicfig} and~\ref{pwtfig} show the comparison of the $P_{SIC}$ at stages $(2,j)$  and $(3,j)$, respectively, for various values of $T$. 
The numerical values of $P_{SIC}(2,j)$ and $P_{SIC}(3,j)$ are obtained using the closed-form expressions derived in (\ref{P_SIC}) and (\ref{P_wt}), respectively.
In general, the $P_{SIC}(i,j)$ for $i=\{2,3\}$ decreases with increasing $j$. 
The reason for this behavior can be explained using the model depicted in Fig.~\ref{Mod12}. From the model, it is observed that the number of devices at state $j$ increases with decreasing value of $j$ as also seen in the analytical expression derived in (\ref{gamma_j}). Subsequently, the chances of success with SIC increases from right to left due to the increase in successful transmissions from other devices and hence the observation. Further, $P_{SIC}$ decreases with increasing value of $T$ as can be observed from Figs.~\ref{psicfig} and~\ref{pwtfig}.

Figs.~\ref{b30_T} and~\ref{b30_K} show the analytical validation of the proposed mechanism in terms of the 
steady-state probability, $b_{4,0}$, of state $s_{4,0}$ in a given cycle 
with respect to $T$ and $K$, respectively.  
The numerical values are obtained using the closed-form expression derived in (\ref{std_b30}). Numerical and simulation results agree well with each other, thus validating our analytical model.
The steady-state probability of success state, $b_{4,0}$, increases linearly with $T$ as shown in Fig.~\ref{b30_T}. This is because that the average number of devices 
that enter the cycle increases with $T$ as given in (\ref{pro_ent}), resulting in more number of successful transmissions. This in turn also improves the number of RACH successes due to SIC.
Further, we observe that for any fixed value of $T$, $b_{4,0}$ increases with $E$ and is maximum at $E=0.6$. Thereafter, it decreases with further increase in $E$. 
For example, for $T=500$, the values of $b_{4,0}$ correspond to $E=0.2$, $0.4$, $0.6$, $0.8$, and $1$ are $0.049$, $0.099$, $0.13$, $0.11$, and $0.098$, respectively. Further details correspond to the effect of $E$ on $b_{4,0}$ are presented in Fig.~\ref{maxe}.
Fig.~\ref{b30_K} shows that $b_{4,0}$ also increases linearly with $K$.
Here, $K$ represents the available orthogonal resources (also called preambles) and the increase in the orthogonal resources increases the number of successful transmissions in a radio frame. This in turn further enhances the number of RACH successes with SIC.
\begin{table}[t]
\begin{center}
\caption{The parameters consider for the performance comparison of different RACH mechanisms using Simulation.}
\label{param}
\begin{tikzpicture}[x=.75cm,y=.5cm][every text node part/.style={align=center}]
\linespread{0.8}
\draw (0,-5) -- (11,-5) -- (11,5) -- (0,5) -- (0,-5);
\draw (0,3) -- (11,3);
\draw (2.5,4) -- (11,4);
\draw (2.5,-5) -- (2.5,5);
\draw (4.5,-5) -- (4.5,4);
\draw (6.5,-5) -- (6.5,4);
\node at (3.5,3.5){EAB$_{(i=1)}$};
\node at (5.5,3.5){FRM$_{(i=2)}$};
\node at (8.75,3.5){Proposed mechanism$_{(i=3)}$};
\node at (1.25,4){Parameter};
\node at (1.25,2.4)[text width = 2cm, align = center]{Total MTC \\ devices};
%\node at (1.5,-0.5) [text width = 2cm]{\scriptsize{MTC} \\ \scriptsize{device}};
\node at (3.5,2.5){$10^5$};
\node at (5.5,2.5){$10^5$};
\node at (8.75,2.5){$10^5$};
%\node at (1.25,1.5){$C$};
%\node at (3.5,1.5){$0.8$};
%\node at (5.5,1.5){$-$};
%\node at (8.25,1.5){$-$};
\node at (1.25,0.5){$A_i$};
\node at (3.5,0.5){$0.8$};
\node at (5.5,0.5){$10^{-5}$};
\node at (8.75,0.5){$10^{-5}$};
\node at (1.25,-0.5){$T$};
\node at (3.5,-0.5){$1482$};
\node at (5.5,-0.5){$-$};
\node at (8.75,-0.5){$1482$};
\node at (1.25,-1.5){$X$};
\node at (3.5,-1.5){$-$};
\node at (5.5,-1.5){$54$};
\node at (8.75,-1.5){$-$};
\node at (1.25,-2.5){$Y$};
\node at (3.5,-2.5){$-$};
\node at (5.5,-2.5){$20$};
\node at (8.75,-2.5){$-$};
\node at (1.25,-3.5){$R$};
\node at (3.5,-3.5){$-$};
\node at (5.5,-3.5){$-$};
\node at (8.75,-3.5){$2-5$};
\node at (1.25,-4.5){$E$};
\node at (3.5,-4.5){$-$};
\node at (5.5,-4.5){$-$};
\node at (8.75,-4.5){$0.6$};
\node at (6.25,4.5){Mechanism};

\end{tikzpicture}
\end{center}
\end{table}

Figs.~\ref{maxe} and \ref{compacc1} show the simulation results to find the optimal values of fractional $E$ and $R$ that maximize the number of RACH successes, respectively, for various values of $T$. From Fig.~\ref{maxe}, we observe that the number of RACH successes are maximum at $E=0.6$. 
For $E<0.6$, the number of devices that enters into the radio frame cycle is less resulting in a limited number of RACH successes and for $E>0.6$ the number of devices that enter into the cycle are more resulting in less number of successes due to more collisions.
Thus, in both cases, the success rate is less. 
The analytical value of optimum $E$ can 
possibly be obtained by maximizing the expression derived in (\ref{std_b30})
with respect to $E$ and is a possible extension of this work. 
From Fig.~\ref{compacc1}, we observe that the number of RACH successes corresponding to $R=1$ (3GPP-EAB mechanism) is very low as there is no SIC. 
In general, it is seen that as $T$ increases  the number of RACH successes increases for all $R$. It is also observed from Fig.~\ref{compacc1} that the number of RACH successes is a concave function of $R$. This can be justified by the fact that for the higher values of $R$, more collisions occur than SIC. Similarly, for the smaller values of $R$, the chances of SIC is very less and it may not contribute much to the success rate.
\begin{figure}[t]
\centering
\epsfig{file=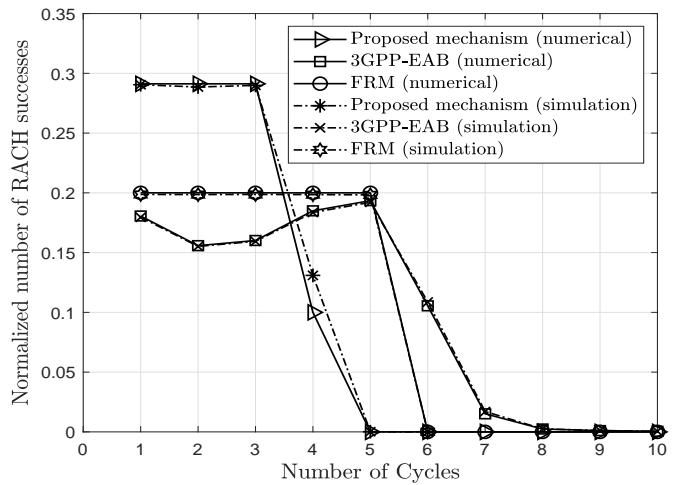, width=3.45 in, height=2.5in}
\caption{Number of RACH successes made in a radio frame cycle for 3GPP-EAB, FRM, and the proposed RACH mechanism with $T=1000$, $K =54$, $R=2$, and $E=0.6$.}
\label{compsucc2ana}
\end{figure}
\begin{figure}[t]
\centering
\epsfig{file=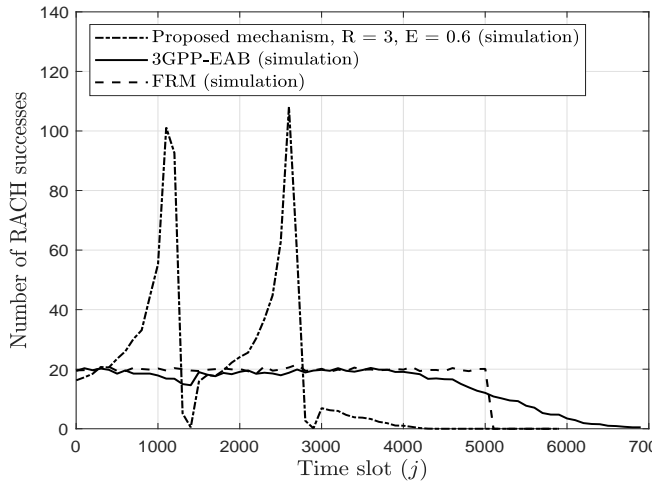, width=3.45 in, height=2.5 in}
\caption{Number of RACH successes made in a radio frame for 3GPP-EAB, FRM, and the proposed RACH mechanism with $T = 1482$, $K =54$, $R=3$, and $E=0.6$.}
\label{compsucc2}
\end{figure}
\subsection{Comparison with the Existing Mechanisms}
In this subsection, we compare the performance of the proposed RACH mechanism with the existing ones. 
The values of all the parameters used in this subsection are listed in Table.~\ref{param}. In the previous subsection, we assumed that at the beginning of every cycle there are 0.1 million devices ready to access the BS. However, in this subsection, we consider a total of 0.1 million devices, distributed over all cycles, for the entire analysis. This constraint is added to compare the efficiency of the different RACH mechanisms with the same number of devices.

Fig.~\ref{compsucc2ana} shows the numerical and simulation 
comparison of the proposed mechanism with the existing ones in terms of the normalized number of RACH successes in each radio frame cycle. 
The numerical results are obtained using the closed-form expression derived in (\ref{std_b30}).
Given that there exists no option of $T$ in the FRM mechanism \cite{dama7}, we obtain the number of RACH successes in each radio frame cycle by
adding up the number of RACH successes in corresponding $T$ radio frames. 
From Fig.~\ref{compsucc2ana}, we observe that to make all $0.1$ million devices successful, the 3GPP-EAB, FRM, and the proposed mechanism require nearly $8$, $5$, and $4$ radio frame cycles, respectively. Thus, we conclude that our proposed  mechanism outperforms the existing ones. This happens because the proposed mechanism allows limited number of repeated transmissions and applies SIC.

Fig.~\ref{compsucc2} shows the 
comparison of the different mechanisms in slot wise manner (unlike the last one that shows radio frame cycle wise). We consider
$R=3$ in the proposed mechanism for a fair comparison.
As shown in~\cite{dama7}, the 3GPP-EAB mechanism allows an average of $54$ devices to contend in a given radio frame resulting an average of $20$ devices to get success. 
However, an MTC device enters the contention loop remain there till it gets success, and a new set of MTC devices enter the loop at the beginning of each cycle.
This results in an increase in the channel contention as shown in~\cite{dama7}.
To mitigate this effect, an FRM mechanism has been proposed in~\cite{dama7} that allows only $54$ devices to contend in a radio frame. This results in a maximum number of $20$ successes in a radio frame. 
Whereas, the proposed SIC based RACH mechanism results in an average of $33$ devices to get success in a radio frame. This increase in the success rate is due to the utilization of back-and-forth SIC at the BS. Further, we conclude that the proposed mechanism outperforms the existing ones as seen in Fig.~\ref{compsucc2}.
\begin{figure}[t]
\centering
\epsfig{file=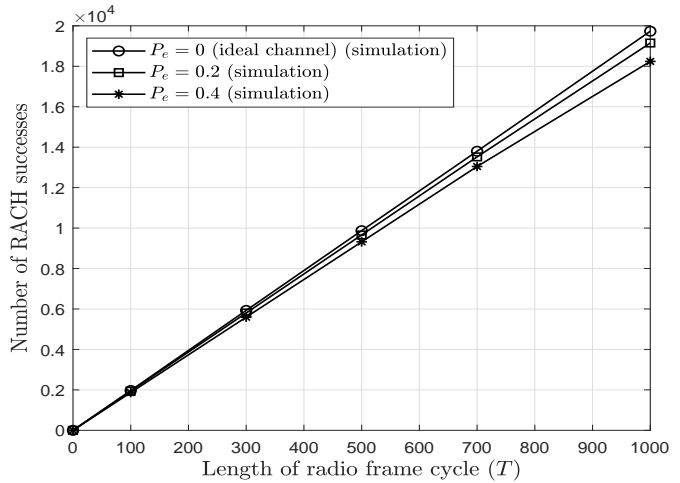, width=3.45 in, height=2.5 in}
\caption{Number of RACH successes made in a radio frame cycle for the proposed RACH mechanism with respect to $T$ for various values of $P_e$ with $K =54$, $R=2$, and $E=1$.}
\label{eff_pe}
\end{figure}

\subsection{The effect of PHY layer impairments on the success rate}
In the previous subsection, the performance of the proposed mechanism is evaluated under ideal PHY channel conditions. In this subsection, we analyse the effect of PHY layer impairments on the performance of the proposed RACH mechanism. In general, the PHY layer impairments, such as additive noise, error in channel estimation, etc., result in bit error which in-turn causes packet error \cite{chawla}. 
However, such errors can be detected and corrected up to some extent by employing error correcting codes but cannot be eliminated completely \cite{han}. 
Hence, in this section, we introduce an arbitrary probability of packet error, denoted by $P_e$, to incorporate the effect of PHY layer impairments in our analysis. 
Here, $P_e$ represents the probability that 
a device is unsuccessful
due to PHY layer impairments even though it alone transmits a unique preamble in a radio frame (successful RCAH).
% However, the error correcting codes are used at the PHY layer to  The packet error is a better metric to evaluate the performance of the network \cite{chawla}. These errors can be corrected using coding schemes whereas all the errors cannot be corrected. Thus, we consider the arbitrary values of probability of error $(P_e)$ representing the probability that a message that is transmitted is corrupted. 

Fig.~\ref{eff_pe} shows the simulation results to find  the effect of $P_e$ on the performance of the proposed SIC based RACH mechanism. We consider arbitrary values of $P_e=0.2$ and $0.4$ for a fair comparison. It is observed that there is a slight decrement in the number of RACH successes with an increase in $P_e$. 
This is due to the fact that PHY layer impairments will reduce the number of devices entering into the contention loop, and hence, the overall success rate.
\section{conclusion} \label{CON}
In this paper, an SIC based RACH mechanism is proposed for a cellular Internet of Things. The primary objective of the proposed mechanism is to improve the over all success rate by employing SIC at the BS. All the devices are allowed to transmit repeatedly for a finite number of times in a given cycle and thereafter back-and-forth SIC is applied at the BS. A novel analytical framework of the proposed mechanism has been developed with all transition and steady-state probabilities for the repetition rate two. Moreover, the analytical expression of the probability of SIC, for a given slot, has been derived in closed-form. Through the extensive numerical results, it has been concluded that the proposed mechanism outperforms the existing 3GPP-EAB and FRM mechanisms in terms of the success rate. Further, to obtain the maximum success rate, the optimum number of devices to be entered in a cycle and the repetition rate have been calculated numerically. In this work, we have developed an analytical framework of the proposed mechanism for the repetition rate two, in future, we will develop an analytical model for a generalized repetition rate.
%\section*{acknowledgment}
%
%This work was supported in part by the Science and Engineering Research Board (SERB), Govt. of India through its Early Career Research (ECR) Award (Ref. No. ECR/2016/001377), and the Department of Science and Technology (DST), Govt. of India (Ref. No. TMD/CERI/BEE/2016/059(G)).

% Can use something like this to put references on a page
% by themselves when using endfloat and the captionsoff option.
\ifCLASSOPTIONcaptionsoff
  \newpage
\fi

% trigger a \newpage just before the given reference
% number - used to balance the columns on the last page
% adjust value as needed - may need to be readjusted if
% the document is modified later
%\IEEEtriggeratref{8}
% The "triggered" command can be changed if desired:
%\IEEEtriggercmd{\enlargethispage{-5in}}

% references section

% can use a bibliography generated by BibTeX as a .bbl file
% BibTeX documentation can be easily obtained at:
% http://mirror.ctan.org/biblio/bibtex/contrib/doc/
% The IEEEtran BibTeX style support page is at:
% http://www.michaelshell.org/tex/ieeetran/bibtex/
%\bibliographystyle{IEEEtran}
% argument is your BibTeX string definitions and bibliography database(s)
%\bibliography{IEEEabrv,../bib/paper}
%
% <OR> manually copy in the resultant .bbl file
% set second argument of \begin to the number of references
% (used to reserve space for the reference number labels box)
\IEEEtriggeratref{14}

\end{document}